\documentclass[journal]{IEEEtran}
\usepackage[dvips]{graphicx}
\usepackage{epsfig}
\usepackage{amsmath}
\usepackage{epsfig}
\usepackage{array}
\usepackage{amssymb}
\usepackage{color}
\newtheorem{Assumption}{\textbf{Assumption}}
\newtheorem{Prop}{\textbf{Property}}
\newtheorem{Lemma}{\textbf{Lemma}}
\newtheorem{Corollary}{\textbf{Corollary}}

\begin{document}

\title{Secure Degree of Freedom of Wireless Networks Using Collaborative Pilots}

\author{Yingbo Hua,
 \emph{Fellow, IEEE}
\thanks{This paper was accepted for publication in IEEE Transactions on Signal Processing. Y. Hua and M. S. Rahman are with Department of Electrical and Computer Engineering,
University of California, Riverside, CA 92521, USA. Q. Liang is with Southwest Jiaotong University, Sichuan, China. Emails: yhua@ece.ucr.edu, qingpengliang@gmail.com, mrahm054@ucr.edu. This work was supported in part by the Army Research Office under Grant Number W911NF-17-1-0581 and the Department of Defense under W911NF-20-2-0267. The views and conclusions contained in this
document are those of the author and should not be interpreted as representing the official policies, either
expressed or implied, of the Army Research Office or the U.S. Government. The U.S. Government is
authorized to reproduce and distribute reprints for Government purposes notwithstanding any copyright
notation herein. Q. Liang is supported by Fundamental Research Funds for the Central Universities under Grant 2682022CX022.}
Qingpeng Liang,
\emph{Member, IEEE},
and Md Saydur Rahman, \emph{Student Member, IEEE}
}

\maketitle

\begin{abstract}
A wireless network of full-duplex nodes/users, using anti-eavesdropping channel estimation (ANECE) based on collaborative pilots, can yield a positive secure degree-of-freedom (SDoF) regardless of the number of antennas an eavesdropper may have.
This paper presents novel results on SDoF of ANECE by analyzing secret-key capacity (SKC) of each pair of nodes in a network of multiple collaborative nodes per channel coherence period.  Each transmission session of ANECE has two phases: phase 1 is used for pilots, and phase 2 is used for random symbols. This results in two parts of SDoF of ANECE. Both lower and upper bounds on the SDoF of ANECE for any number of users are shown, and the conditions for the two bounds to meet are given. This leads to important discoveries, including: a) The phase-1 SDoF is the same for both multi-user ANECE and pair-wise ANECE while the former may require only a fraction of the number of time slots needed by the latter; b) For a three-user network, the phase-2 SDoF of all-user ANECE is generally larger than that of pair-wise ANECE; c) For a two-user network, a modified ANECE deploying square-shaped nonsingular pilot matrices yields a higher total SDoF than the original ANECE. The multi-user ANECE and the modified two-user ANECE shown in this paper appear to be the best full-duplex schemes known today in terms of SDoF subject to each node using a given number of antennas for both transmitting and receiving.
\end{abstract}


\section{Introduction}
Full-duplex radio wireless networks have attracted much attention in recent years. The feasibility of full-duplex has been enabled by both signal processing algorithms and microwave circuit implementations, e.g., see \cite{Kolodziej2019} and \cite{Hua2015}. It is widely known that full-duplex can be used to enhance both the spectral efficiency of a radio network and the security of the radio network against eavesdropping. This paper is concerned with the latter.

In the field of wireless security, physical layer security is an important topic because of its ability to provide information-theoretic security between a pair of nodes or users even in the absence of a pre-existing secret key \cite{Poor2017}. (We will use ``node'' and ``user'' as interchangeable terms.) However, numerous studies show that with only half-duplex radio nodes, the secrecy between the nodes goes to zero if the channel between the nodes lacks any reciprocal property and the eavesdropping channel is stronger than the legitimate channel, e.g., see \cite{Khisti2010} and \cite{Liu2015}. A strong eavesdropping channel may be due to an eavesdropper (Eve) having more antennas than either the legitimate transmitter (Alice) or the legitimate receiver (Bob), or simply due to Alice being closer to Eve than to Bob.

With full-duplex radio coming over the horizon of future wireless networks, many researchers have explored the advantages of full-duplex for wireless security in a variety of settings \cite{Zheng2013}-\cite{Zabir2020}. For example, a full-duplex destination node can receive a signal from a source node, and at the same time and same frequency transmit a jamming noise to reduce the signal-to-noise ratio (SNR) at Eve. Also, for example, a full-duplex multi-antenna relay can receive a signal from a source node, relay another signal to a destination node and transmit a jamming noise to interfere Eve, all at the same time and same frequency. The literature shows various problem formulations and corresponding solutions for power optimization at source, destination and/or relay to maximize a secrecy capacity subject to some partial or full knowledge of Eve's receive channel.

Full-duplex radio is also considered in \cite{Li2016}-\cite{Li2018}, where secure degrees of freedom (SDoF) of their proposed schemes were analyzed. The work in \cite{Mahmood2017} only considers a single antenna on each user and Eve. The SDoF of the schemes in \cite{Li2016} and \cite{Li2018} reduce to zero when the number of antennas on Eve is no less than the sum of the transmit antennas at Alice and Bob. Those findings have significantly underestimated the potentials of full-duplex in terms of SDoF, which will become clear later.

We like to mention that an early study of SDoF was shown in \cite{He2014} and later in \cite{Liu2015} for a half-duplex MIMO channel-model (CM) based secrecy capacity, where SDoF is shown to be zero if and only if the number of antennas on either Alice or Bob is no larger than that on Eve.
More recently, SDoF is analyzed in \cite{Hua2023} for a half-duplex MIMO source-model (SM) based secrecy capacity (also known as secret-key capacity), where (assuming imperfect channel reciprocity) SDoF is shown to be zero if and only if the number of antennas on each of Alice and Bob is no larger than that on Eve.

It appeared not until the work \cite{Hua2019} that researchers started to pay much attention to the impact of Eve who may have not only unlimited computing power but also a large number of antennas. The work \cite{Hua2019} discovered a method that allows transmission of secret information between two or more full-duplex nodes regardless of the number of antennas on Eve even if there is no reciprocal property in the channels between the legitimate nodes.
The method shown in \cite{Hua2019} is called anti-eavesdropping channel estimation (ANECE), which uses collaborative pilots to exploit the condition that at the beginning of every channel coherence period, no node knows its receive channel state, and  channel estimation (including carrier synchronization) must be conducted for any reliable transmission between users.
ANECE is based on full-duplex, which is unlike the discriminatory channel estimation (DCE) methods based on half-duplex \cite{Chang2010}, \cite{Huang2013}, \cite{Yang2014} and \cite{Wang2015}. Without a channel reciprocal property between Alice and Bob, DCE is unable to yield a positive secrecy if the number of antennas on Eve is relatively large.

Further developments and understandings of ANECE are available in \cite{Sohrabi2019}, \cite{Zhu2020} and \cite{WuHua2022}. Lower bounds on the degree of freedom (DoF) of secrecy capacity for a two-user ANECE were first reported in \cite{Sohrabi2019}. Optimal designs of the pilots required by a multi-user ANECE were developed in \cite{Zhu2020}. A total secrecy of ANECE in both channel estimation phase (phase 1) and random symbol transmission phase (phase 2) was studied in \cite{WuHua2022}. It was shown in \cite{WuHua2022} that a lower bound on the SDoF of ANECE is positive even if there is no channel reciprocity and Eve has an unlimited number of antennas. Furthermore, a related study of SDoF of two full-duplex nodes without initial knowledge of channel state information was studied in \cite{Liang2020}.
Despite the prior works on ANECE, many important questions remained. For example, the work in \cite{Liang2020} only handles two full-duplex nodes instead of a network of more than two full-duplex nodes in each channel coherence period. The SDoF results shown in \cite{WuHua2022} are based on the optimal pilots designed in  \cite{Zhu2020}, instead of the most basic constraints on the collaborative pilots needed for ANECE.

This paper aims to reveal further insights into SDoF of ANECE. We will adopt secret-key capacity (SKC) as a generalized secrecy capacity for both phases of ANECE. SKC is also known as the SM based secrecy capacity. Unlike the CM based secrecy capacity which allows public communication (for sharing a coding scheme for example) only before the transmission of a secret, SKC allows public communication both before and after the transmission of a secret \cite{Maurer1993}. While the ``secret'' transmitted using CM and the ``secret'' extracted from SM generally differ in ``content'',  the secrecy in both cases can be measured in the same way if we ignore the cost and timing of operations. For example, any secret key of $n$ bits extracted using SM can be used to keep any information of $n$ bits in total secrecy even via public communications. So, it is meaningful to treat SKC as a generalization of secrecy capacity. In fact, a similar notion was used for a half-duplex SISO channel in \cite{Lai2012} and recently for a half-duplex MIMO channel in \cite{Hua2023}. In this paper, we will use SKC and secrecy capacity interchangeably unless mentioned otherwise.

More specifically, we consider $M$ multi-antenna (full-duplex) nodes/users cooperatively performing the two-phased ANECE within each channel coherence period $\mathcal{P}$ against a multi-antenna Eve. This is also referred to as multi-user ANECE. We aim to determine its pair-wise SDoF defined as $\texttt{SDoF}_{i,j}\doteq\texttt{DoF}(C_{\texttt{key}, i,j})\doteq\lim_{\sigma^2\to\infty}\frac{C_{\texttt{key}, i,j}}{\log\sigma^2}$, where $C_{\texttt{key}, i,j}$ is the SKC (in bits per $\mathcal{P}$) between node $i$ and node $j$ based on the observations from the execution of multi-user ANECE, and $\sigma^2$ is a nominal transmission power consumed by each node for multi-user ANECE. It will become clear that  $\texttt{SDoF}_{i,j}=\texttt{SDoF}_{i,j}^{(1)}+\texttt{SDoF}_{i,j}^{(2)}$ with $\texttt{SDoF}_{i,j}^{(1)}$ and $\texttt{SDoF}_{i,j}^{(2)}$ corresponding to phases 1 and 2 of multi-user ANECE. We will also study $\texttt{SDoF}_{i,j}$ for a pair-wise ANECE where each $\mathcal{P}$ is shared orthogonally by all pairs of users among $M$ users and each pair performs a two-user ANECE. Furthermore, we will study a modified two-user ANECE for a two-user network (i.e., only two users are considered within each $\mathcal{P}$). Different coherence periods are assumed to be independent.

Major results in this paper are summarized in the properties at the end of each relevant section. The most significant ones are shown in Properties \ref{Property_major}-\ref{Property_two_users_comparison}. In particular, we will show:

1) Assuming reciprocal channels between users and a full rank condition on pilots, $\texttt{SDoF}_{i,j}^{(1)}=N_iN_j$ for both multi-user ANECE and pair-wise ANECE with $N_i$ denoting the number of antennas on user $i$. This holds for all $M\geq 2$ and $N_E\geq 0$ with $N_E$ denoting the number of antennas on Eve. But the pair-wise ANECE requires $\frac{M}{2}$ times more time-slots for phase-1 transmissions.

2) For a three-user symmetric network where $M=3$ and $N_i=N$ for all $i$,  $\texttt{SDoF}_{i,j}^{(2)}=0$ for pair-wise ANECE if $N_E\geq 2N$ while $\texttt{SDoF}_{i,j}^{(2)}>0$ for multi-user ANECE if $1\leq K_2<2N$ and (even)  $N_E\geq 3N$. Here $K_2$ is the number of time-slots used for phase-2 transmissions.

  3) For a two-user network where $M=2$ and $N_1\leq N_2$, $\texttt{SDoF}_{1,2}^{(2)}=2N_1^2$ for the original two-user ANECE (and also for the scheme in  \cite{Liang2020}) if $N_E\geq N_1+N_2$ and $K\geq N_1+N_2$, but $\texttt{SDoF}_{1,2}^{(2)}=N_1(N_1+N_2)$ for the modified two-user ANECE under the same conditions. Here $K$ is the total number of time-slots used for both phase-1 and phase-2 transmissions.

 These results are novel and significant. (The notations used later for SDoFs will be slightly different for better referencing.)

The rest of the paper is organized as follows.
In section \ref{sec:Preliminary}, we review the meanings of DoF and SDoF, and introduce several lemmas that are important throughout this paper. The notion of pair-wise SKC for multi-user ANECE is defined.
In section \ref{sec:multiuser_ANECE}, we show the key features of multi-user ANECE, including the structure of the collaborative pilots and random symbols used by multi-user ANECE. The effects of the collaborative pilots on channel estimation at users and Eve are highlighted.
In section \ref{sec:decomposition}, we show how the pair-wise SKC of multi-user ANECE can be decomposed into two components: one for phase 1 and the other for phase 2. These two components will be treated separately in the sequel of the paper.
In section \ref{sec:Phase_1}, we derive the pair-wise SDoF for phase 1 of multi-user ANECE subject to reciprocal random channel parameters between users, which leads to a result more general than one in \cite{WuHua2022}.
In section \ref{sec:Phase_2}, the pair-wise SDoF for phase 2 of multi-user ANECE is shown in terms of its lower and upper bounds. The conditions for the bounds to meet are also given.
In section \ref{sec:Pairwise}, we compare all-user ANECE with a pair-wise ANECE for a network of $M\geq 3$ users. We show that all-user ANECE has advantages over pair-wise ANECE in both phases 1 and 2.
In section \ref{sec:general_ANECE}, we propose and analyse a modified two-user ANECE for a two-user network, for which each user applies a square-shaped pilot matrix.
In section \ref{sec:Comparison}, we show that the SDoF of the modified two-user ANECE is generally larger than that of the original two-user ANECE.
Section \ref{sec:conclusion} concludes the paper.

\emph{Notations:} The superscripts $^T$, $^*$ and $^H$ denote transpose, conjugate and conjugate transpose respectively while $*$ as an entry in a matrix denotes a quantity whose specific form is not important for discussion. $\mathbb{C}^{m\times n}$ denotes the space of $m\times n$ complex matrices. $\mathbb{E}$, $\otimes$ and $\oplus$ are respectively expectation, Kronecker product and (element-wise) exclusive-OR. $(a)^+=\max(a,0)$. $|\mathbf{A}|$ is the determinant of the matrix $\mathbf{A}$. All matrices and vectors are represented respectively by boldface upper cases and boldface lower cases. $\doteq$ stresses ``equal by definition''. $\mathcal{CN}(\mathbf{m},\mathbf{R})$ denotes the circular complex Gaussian probability-density-function (PDF) with mean vector $\mathbf{m}$ and covariance matrix $\mathbf{R}$.  $\mathbb{I}(A;B|C)$ denotes the mutual information between $A$ and $B$, conditional on $C$. $h(A|B)$ is the differential entropy of $A$ conditional on $B$ (or entropy of $A$ conditional on $B$ if $A$ is discrete). And $(\cdot)$, $[\cdot]$ and $\{\cdot \}$ are used interchangeably.

\section{Preliminaries}\label{sec:Preliminary}

\subsection{Degree of Freedom and Secure Degree of Freedom}
The degree of freedom (DoF) of
a function $g(\sigma^2)$ of $\sigma^2$ is said to be $d$ relative to $\log\sigma^2$ if for large $\sigma^2$, $g(\sigma^2)\approx d\log\sigma^2+c$ where both $d$ and $c$ are invariant to $\sigma^2$. In this case, we also write $\texttt{DoF}(g(\sigma^2))=d$. The approximation ``$\approx$'' we use in this paper will be always such that it does not affect the DoF.
If a scheme has the secrecy capacity $C_s$ and $\texttt{DoF}(C_s)=d$, we say that this scheme has the secure degree of freedom (SDoF) equal to $d$.

The following two lemmas and corollary will be used frequently.
\begin{Lemma}\label{Lemma1}
Let $\mathbf{R}$ be an $n\times n$ complex positive-definite matrix dependent on $\sigma^2$. If for large $\sigma^2$  the $i$th eigenvalue $\lambda_i$ of $\mathbf{R}$ can be written as $\eta_i\sigma^2+a_i$ for $1\leq i\leq m$, and as $b_i$ for $m+1\leq i\leq n$, where $\eta_i>0$, $a_i$ and $b_i>0$ are all invariant to $\sigma^2$, then  $\texttt{DoF}(\log_2 |\mathbf{R}|)=m$.
\end{Lemma}
\begin{IEEEproof}
It is known that
$\log_2|\mathbf{R}|=\log_2\prod_{i=1}^n\lambda_i =\sum_{i=1}^n \log_2\lambda_i$. As $\sigma^2$ increases, $\log_2|\mathbf{R}|$ converges to $\sum_{i=1}^m\log_2 (\eta_i \sigma^2+a_i)+\sum_{i=m+1}^n \log b_i = m\log_2\sigma^2+\sum_{i=1}^m\log_2(1+\frac{a_i}{\eta_i\sigma^2})+\sum_{i=1}^m \log_2\eta_i+\sum_{i=m+1}^n \log_2 b_i$ which further converges to $ m\log_2\sigma^2+\sum_{i=1}^m \log_2\eta_i+\sum_{i=m+1}^n \log_2 b_i$. Hence $\texttt{DoF}(\log_2 |\mathbf{R}|)=m$.
\end{IEEEproof}

\begin{Corollary}\label{Corollary1} Let $\mathbf{R}$ be a (continuous) random covariance matrix.
  If  with probability one $\mathbf{R}$ converges to $ \sigma^2\mathbf{R}_m+\mathbf{I}_n$ as $\sigma^2$ increases to $\infty$ where $\mathbf{R}_m$ has the rank $m\leq n$, then  $\texttt{DoF}(\mathbb{E}\{\log_2 |\mathbf{R}|\})=m$.
\end{Corollary}
\begin{IEEEproof}
This follows from Lemma \ref{Lemma1}.
\end{IEEEproof}

\begin{Lemma}\label{Lemma:DoF_Gaussian}
Let $\mathbf{Y}=\sigma\mathbf{H}\mathbf{X}+\mathbf{W}\in \mathbb{C}^{m\times k}$ where $\mathbf{H}\in\mathbb{C}^{m\times n}$, $\mathbf{X}\in\mathbb{C}^{n\times k}$ and $\mathbf{W}\in \mathbb{C}^{m\times k}$ are independent of each other and consist independent and identically distributed (i.i.d.) $\mathcal{CN}(0,1)$ entries. Then, relative to $\log \sigma^2$,
$\texttt{DoF}(h(\mathbf{Y}|\mathbf{H}))=\min(m,n)k$ and $\texttt{DoF}(h(\mathbf{Y}|\mathbf{X}))=\min(n,k)m$.
\end{Lemma}
\begin{IEEEproof}
Let $\mathbf{y}=vec(\mathbf{Y})$, $\mathbf{x}=vec(\mathbf{X})$ and $\mathbf{w}=vec(\mathbf{W})$. Then $\mathbf{y}=\sigma(\mathbf{I}_k\otimes \mathbf{H})\mathbf{x}+\mathbf{w}$, which has the conditional PDF $f(\mathbf{y}|\mathbf{H})=\mathcal{CN}(0,\mathbf{R})$ with $\mathbf{R}=\sigma^2(\mathbf{I}_k\otimes \mathbf{H}\mathbf{H}^H)+\mathbf{I}_{mk}$. It follows that $h(\mathbf{Y}|\mathbf{H})=\mathbb{E}\{\log\frac{1}{f(\mathbf{y}|\mathbf{H})}\}
=\mathbb{E}\{\log(\pi^{mk}|\mathbf{R}|\exp(\mathbf{y}^H\mathbf{R}^{-1}\mathbf{y}))\}
=\mathbb{E}\{\log(e^{mk}\pi^{mk}|\mathbf{R}|)\}$. It follows from Corollary \ref{Corollary1} that $\texttt{DoF}(h(\mathbf{Y}|\mathbf{H}))=\min(m,n)k$. Also note $\texttt{DoF}(h(\mathbf{Y}|\mathbf{X}))=\texttt{DoF}(h(\mathbf{Y}^T|\mathbf{X}))=\min(k,n)m$.
\end{IEEEproof}

The next lemma will be useful for cases where the Gaussian assumption in Lemma \ref{Lemma:DoF_Gaussian} is not available.
\begin{Lemma}\label{Lemma:DoF_nonGaussian}
Let $\mathbf{y}=\sigma \mathbf{x}+\mathbf{w}\in \mathbb{C}^{n\times 1}$ where $\mathbf{w}$ is noise and all entries in $\mathbf{x}$ can be asymptotically determined by minimal $d$ entries of $\mathbf{y}$ as $\sigma^2$ increases. Then relative to $\log\sigma^2$, $\texttt{DoF}(h(\mathbf{y}))=d$ (regardless of the PDF of $\mathbf{y}$).
\end{Lemma}
\begin{IEEEproof}
For any real-valued random variable $X$ with a PDF $f_X(x)$, it is known that another random variable $Y=\sigma X$ has the PDF $f_Y(y) = \frac{1}{\sigma}f_X(\frac{y}{\sigma})$. Then $h(Y)=\int f_Y(y)\log\frac{1}{f_Y(y)}dy=\log \sigma +h(X)$. For a complex-valued vector $\mathbf{y}$, we now assume that its first $d$ (complex) elements are the minimum number of elements needed to determine all elements in $\mathbf{x}$ at high power $\sigma^2$. Then we have $h(\mathbf{y})=h(y_1)+h(y_2|y_1)+\cdots+h(y_d|y_{d-1},\cdots,y_1)+c$ where $y_i$ is the $i$th element of $\mathbf{y}$, and $c=h(y_{d+1}|y_d,\cdots,y_1)+\cdots+h(y_n|y_{n-1},\cdots,y_1)$ which is invariant to $\sigma^2$ for large $\sigma^2$. Note that since $x_{d+m}$ can be determined by $y_1,\cdots,y_d$ at large $\sigma^2$, then as $\sigma^2$ increases, $h(y_{d+m}|y_1,\cdots,y_d)$ converges to $h(w_{d+m}|y_1,\cdots,y_d)$ which is invariant to $\sigma^2$.
On the other hand, for $i\leq d$, since  no $x_i$ (real or imaginary part) can be determined by $y_{i-1},\cdots,y_1$ for large $\sigma^2$,
it follows that $h(y_i|y_{i-1},\cdots,y_1)$ converges to $h(\sigma e_i)$ for large $\sigma^2$ where $e_i$ is the nonzero estimation error of $x_i$ at $\sigma^2=\infty$. Furthermore, $h(\sigma e_i)=2\log\sigma +h(\Re(e_i))+h(\Im(e_i)|\Re(e_i))$.
 Hence, $\texttt{DoF}(h(\mathbf{y}))=d$.
\end{IEEEproof}

\subsection{Pair-Wise Secret-Key Capacities}
Consider $M$ users where user $i$ has the random data set $\mathcal{Y}_i$ with $i=1,\cdots,M$, and Eve who has the random data set $\mathcal{Y}_E$. We will be interested in the secret-key capacity (SKC) of each pair of users against Eve, which is also called pair-wise SKC.

\begin{Lemma}\label{Lemma_key_capacity}
The pair-wise SKC between user $i$ and user $j$ for all $i\neq j$ against Eve in bits per \emph{independent realization} of $\{\mathcal{Y}_1,\cdots,\mathcal{Y}_M,\mathcal{Y}_E\}$, denoted by $C_{\texttt{key},i,j}$, satisfies
\begin{align}\label{eq:Lemma_key_capacity}
 & C_{\texttt{key},i,j}\geq \max(C_{\texttt{low},i,j},C_{\texttt{low},j,i})
\end{align}
with $C_{\texttt{low},i,j}=\mathbb{I}(\mathcal{Y}_i;\mathcal{Y}_j)-\mathbb{I}(\mathcal{Y}_i;\mathcal{Y}_E)$
 and
\begin{equation}\label{eq:Lemma_key_capacity_2}
  C_{\texttt{key},i,j}\leq C_{\texttt{up},i,j}\doteq \mathbb{I}(\mathcal{Y}_i;\mathcal{Y}_j|\mathcal{Y}_E).
\end{equation}
\end{Lemma}
\begin{IEEEproof}
The following proof follows the same principle as in section 4.2.1 in \cite{Bloch2011}.
First we assume that all entries in $\mathcal{Y}_1$, $\cdots$, $\mathcal{Y}_M$ and $\mathcal{Y}_E$ are discrete (after quantization), or equivalently each of the sets $\mathcal{Y}_1$, $\cdots$, $\mathcal{Y}_M$ and $\mathcal{Y}_E$ becomes a finite multidimensional discrete symbol belonging to a common set $\mathbb{Y}$. For each $i$, user $i$ generates a uniformly random symbol $\mathcal{U}_i$ from $\mathbb{Y}$ and transmits $\mathcal{\bar Y}_i=\mathcal{Y}_i\oplus\mathcal{U}_i$ via public channel. These transmissions constitute $M$ broadcast wiretap channels, i.e., from each user to all other users. It follows that the secrecy capacity from user $i$ to user $j$ against Eve in this case is
\begin{align}\label{eq:lemma_key_1}
  &
\mathbb{I}[\mathcal{U}_i;\mathcal{Y}_j,\mathcal{\bar Y}_1,\cdots,\mathcal{\bar Y}_M]-
\mathbb{I}[\mathcal{U}_i;\mathcal{Y}_E,\mathcal{\bar Y}_1,\cdots,\mathcal{\bar Y}_M]\notag\\
&=h[\mathcal{U}_i|\mathcal{Y}_E,\mathcal{\bar Y}_1,\cdots,\mathcal{\bar Y}_M]-
h[\mathcal{U}_i|\mathcal{Y}_j,\mathcal{\bar Y}_1,\cdots,\mathcal{\bar Y}_M]
\end{align}
Since the uniformly random $\mathcal{U}_i$ is independent of $\mathcal{\bar Y}_j$ for all $j\neq i$, and $\mathcal{\bar Y}_j$ is independent of $\mathcal{Y}_E$, $\mathcal{Y}_i$ and $\mathcal{\bar Y}_i$ for all $i\neq j$, \eqref{eq:lemma_key_1} is equivalent to
\begin{align}\label{eq:lemma_key_2}
  &
h[\mathcal{U}_i|\mathcal{Y}_E,\mathcal{\bar Y}_i]-
h[\mathcal{U}_i|\mathcal{Y}_j,\mathcal{\bar Y}_i]=h[\mathcal{Y}_i|\mathcal{Y}_E]-
h[\mathcal{Y}_i|\mathcal{Y}_j]\notag\\
&=\mathbb{I}(\mathcal{Y}_i;\mathcal{Y}_j)-\mathbb{I}(\mathcal{Y}_i;\mathcal{Y}_E)
\end{align}
where we have used $h[\mathcal{U}_i|\mathcal{Y}_E,\mathcal{\bar Y}_i]=h[\mathcal{U}_i,\mathcal{\bar Y}_i|\mathcal{Y}_E]-h[\mathcal{\bar Y}_i|\mathcal{Y}_E]=h[\mathcal{U}_i|\mathcal{Y}_E]+h[\mathcal{\bar Y}_i|\mathcal{U}_i,\mathcal{Y}_E]-h[\mathcal{\bar Y}_i|\mathcal{Y}_E]=h[\mathcal{\bar Y}_i|\mathcal{U}_i,\mathcal{Y}_E]=h[\mathcal{Y}_i|\mathcal{Y}_E]$.
Since $C_{\texttt{key},i,j}=C_{\texttt{key},j,i}$, the above result leads to  \eqref{eq:Lemma_key_capacity}. For continuous $\mathcal{Y}_1,\cdots,\mathcal{Y}_M$ and $\mathcal{Y}_E$, \eqref{eq:Lemma_key_capacity} then follows from the generalized definition of mutual information as shown in \cite{Cover2006}. Finally, \eqref{eq:Lemma_key_capacity_2} follows directly from the upper bound for the two-user case as shown in \cite{Maurer1993} and \cite{Bloch2011}.
\end{IEEEproof}

Note that a key generated by a pair of users may be correlated with a key generated by another pair of users. Even $\mathcal{U}_i$ and $\mathcal{U}_j$ become correlated when conditioned on $\mathcal{\bar Y}_i$ and $\mathcal{\bar Y}_j$. For example, $\mathbb{I}(\mathcal{U}_i;\mathcal{U}_j|\mathcal{\bar Y}_i,\mathcal{\bar Y}_j)=
h(\mathcal{U}_i|\mathcal{\bar Y}_i,\mathcal{\bar Y}_j)-h(\mathcal{U}_i|\mathcal{U}_j,\mathcal{\bar Y}_i,\mathcal{\bar Y}_j)
=h(\mathcal{U}_i|\mathcal{\bar Y}_i)-h(\mathcal{U}_i|\mathcal{Y}_j,\mathcal{\bar Y}_i)
=h(\mathcal{Y}_i)-h(\mathcal{Y}_i|\mathcal{Y}_j)=\mathbb{I}(\mathcal{Y}_i;\mathcal{Y}_j)$.
However, as discussed later, some components in the pair-wise secret keys are independent of each other.

\subsection{Secret-key capacity versus secrecy capacity}
For wiretap channel model (a notion for CM based secrecy) where a secret information is directly transmitted from a source to a destination, there is what is commonly called secrecy capacity. However, given a secret key $\mathcal{K}$ of $n$ bits between the source and the destination, the source can send a secret information $\mathcal{I}$ of $n$ bits by sending $\mathcal{I}\oplus\mathcal{K}$ via public channel, and this $\mathcal{I}$ is then received by the destination in complete  secrecy. In other words, a secret-key capacity can be directly translated into a secrecy capacity if additional operation or communication is allowed. For this reason, we treat ``secret-key capacity'' and ``secrecy capacity'' interchangeably unless mentioned otherwise.

\section{Data Models of Multi-User ANECE Using Collaborative Pilots}\label{sec:multiuser_ANECE}

We consider a network of $M$ cooperative full-duplex nodes against Eve in \emph{each channel coherence period}. All these full-duplex nodes are assumed to be ideal full-duplex. The power of the residual self-interference of a practical full-duplex radio is in general proportional to its transmitted power. A practical full-duplex radio can be modelled as an ideal full-duplex radio as long as the transmitted power is no larger than a threshold. This threshold is governed by a target noise level and a residual self-interference power gain, which certainly affects the achievable distance of full-duplex communication. In order for the SDoF theory shown in this paper to be applicable in a given application, that threshold needs to be relatively large. For implementation issues of full-duplex, see \cite{Kolodziej2019}.

The numbers of antennas on the $M$ full-duplex nodes and Eve are denoted by $N_1, N_2, \cdots, N_M$ and $N_E$ respectively. To simplify many of the mathematical expressions, we assume that node $i$ has the same number $N_i$ of transmit antennas as receive antennas. Note that Eve here could represent multiple colluding eavesdroppers with their combined number of antennas equal to $N_E$.

The multi-user ANECE is such that every node transmits its packet (on the same carrier frequency) concurrently with all other nodes. The required precision for the concurrence is only at the symbol level. Each packet within a channel coherence period consists of a pilot matrix and a random symbol matrix, e.g., node $i$ transmits the $N_i\times K_1$ pilot matrix $\mathbf{P}_i$ in phase 1 and the $N_i\times K_2$ symbol matrix $\mathbf{X}_i$ in phase 2. The pilot matrix and the symbol matrix transmitted from the same node can and should be transmitted sequentially in synch with each other up to the carrier frequency level. So, an estimated channel matrix based on a pilot matrix from node $j$ can be used for coherent detection of the symbol matrix from node $j$. But for the purpose of this paper, we do not need any explicit coherent detection at each node.

Define
$\mathbf{P}=\left [ \mathbf{P}_1^H,\cdots, \mathbf{P}_M^H
  \right ]^H$ and $\mathbf{X}=\left [ \mathbf{X}_1^H,\cdots, \mathbf{X}_M^H
  \right ]^H$. Hence, the transmitted signals from all nodes can be expressed as
\begin{equation}\label{eq:PX}
[\mathbf{P},\mathbf{X}]=\left [
\begin{array}{cc}
  \mathbf{P}_1, & \mathbf{X}_1 \\
  \cdots, & \cdots \\
  \mathbf{P}_M, & \mathbf{X}_M
\end{array}
\right ]\in\mathbb{C}^{N_T\times (K_1+K_2)}
\end{equation}
where $N_T\doteq\sum_{i=1}^M N_i$, each row of \eqref{eq:PX} corresponds to an antenna among the nodes, and each column of \eqref{eq:PX} corresponds to a sampling interval.
Furthermore, we let $\mathbf{P}_{(i)}$ be $\mathbf{P}$ without $\mathbf{P}_i$.

\begin{Assumption}\label{Assumption_Pilot}
The collaborative pilot matrices for all-user ANECE (with total $M$ users) meet the following conditions: $\texttt{rank}(\mathbf{P}_i)=N_i$, $\texttt{rank}(\mathbf{P}_{(i)}) = N_T-N_i$ and $\texttt{rank}(\mathbf{P}) = \max_i \texttt{rank}(\mathbf{P}_{(i)})=N_T-N_{\texttt{min}}$
where $N_{\texttt{min}}\doteq\min_i N_i$. Consequently, we need $K_1\geq N_T-N_{\texttt{min}}$.
\end{Assumption}

A simple way to construct such a $\mathbf{P}$ is to first form a full-rank $N_T\times N_T$ matrix and then remove any $N_{\texttt{min}}$ columns.

\begin{Assumption}\label{Assumption_X}
   All entries in $\mathbf{X}$ are i.i.d. $\mathcal{CN}(0,1)$.
 \end{Assumption}

Assumption \ref{Assumption_Pilot} is the most basic requirement on the collaborative pilots to achieve the maximum ambiguity of channel estimation at Eve, which was originally proposed in Appendix E in \cite{Hua2019}.

In phase 1, the signal received by node $i$ is
\begin{equation}\label{eq:Yi1}
  \mathbf{Y}_i^{(1)} = \sigma\sum_{j\neq i} \mathbf{H}_{i,j}\mathbf{P}_j+\mathbf{W}_i^{(1)}
  =\sigma\mathbf{H}_i\mathbf{P}_{(i)}+\mathbf{W}_i^{(1)}
\end{equation}
where $\mathbf{H}_{i,j}$ is the $N_i\times N_j$ channel matrix from node $j$ to node $i$, and $\mathbf{H}_i$ is the $N_i\times (N_T-N_i)$ matrix from the horizontal stack of $\mathbf{H}_{i,j}$ for all $j\neq i$. At a high SNR or a high power $\sigma^2$ in phase 1, node $i$ for every $i$ can estimate $\mathbf{H}_i$ reliably. SNR in phase 1 is the ratio of the pilot power over the noise power.

Also in phase 1, Eve receives
\begin{equation}\label{eq:mYE1}
  \mathbf{Y}_E^{(1)} = \sigma\sum_{i=1}^M \mathbf{H}_{E,i}\mathbf{P}_i+\mathbf{W}_E^{(1)}
  =\sigma\mathbf{H}_E\mathbf{P}+\mathbf{W}_E^{(1)}
\end{equation}
where $\mathbf{H}_{E,i}$ is the channel matrix from node $i$ to Eve, and $\mathbf{H}_E=[\mathbf{H}_{E,1},\cdots,\mathbf{H}_{E,M}]$. Since $\mathbf{P}$ does not have a full row rank, Eve is unable to obtain a consistent estimate of $\mathbf{H}_E$. But she is able to find a part of $\mathbf{H}_E$ as explain below.

Since the $N_T\times K_1$ matrix $\mathbf{P}$ has the rank $N_T-N_{\texttt{min}}$, there exist a $N_T\times (N_T-N_{\texttt{min}})$ orthonormal matrix $\mathbf{Q}_P$ and a $(N_T-N_{\texttt{min}})\times K_1$ matrix $\mathbf{R}_P$ (of rank $N_T-N_{\texttt{min}}$) such that $\mathbf{P}=\mathbf{Q}_P\mathbf{R}_P$. This follows from the standard QR decomposition. There is also a $N_T\times N_{\texttt{min}}$ orthonormal complement matrix $\mathbf{Q}_{P,\perp}$ such that $\mathbf{Q}\doteq [\mathbf{Q}_P,\mathbf{Q}_{P,\perp}]$ is unitary. Let $\mathbf{H}_E' \doteq \mathbf{H}_E\mathbf{Q}=[\mathbf{H}_{E,P},\mathbf{H}_{E,P,\perp}]$ where
$\mathbf{H}_{E,P}=\mathbf{H}_E\mathbf{Q}_P$
and $\mathbf{H}_{E,P,\perp}=\mathbf{H}_E\mathbf{Q}_{P,\perp}$. Then, \eqref{eq:mYE1} becomes
\begin{align}\label{eq:YE1aa}
  \mathbf{Y}_E^{(1)} &= \sigma\mathbf{H}_E\mathbf{Q}\mathbf{Q}^H\mathbf{P}+\mathbf{W}_E^{(1)}
  =\sigma\mathbf{H}_{E,P}\mathbf{R}_P+\mathbf{W}_E^{(1)}.
\end{align}
So at a high SNR (or high power), Eve is able to estimate $\mathbf{H}_{E,P}$ reliably.

Note that given $\mathbf{Y}_E^{(1)}$ at high power, we know that $\mathbf{H}_E\approx \mathbf{\bar H}_E+\mathbf{T}_E\mathbf{N}_E$ where $\mathbf{\bar H}_E$ is the minimum-norm solution of $\mathbf{H}_E$ to $\mathbf{Y}_E^{(1)}
  =\sigma\mathbf{H}_E\mathbf{P}$, the row span of $\mathbf{N}_E$ is the left null space of $\mathbf{P}$, and $\mathbf{T}_E\in\mathbb{C}^{N_E\times N_{\texttt{min}}}$ is ``arbitrary''. But $\mathbf{T}_E$ obtained from $\mathbf{Y}_E^{(1)}$ along with the prior i.i.d. statistics of $\mathbf{H}_E$ is not fully arbitrary. Yet, no entry of $\mathbf{T}_E$ can be consistently estimated from other entries of $\mathbf{T}_E$ when only $\mathbf{Y}_E^{(1)}$ is given. In this case, the entries of $\mathbf{T}_E$ are said to have full freedoms or simply called ``arbitrary''. The connection between $\mathbf{T}_E\in\mathbb{C}^{N_E\times N_{\texttt{min}}}$ and $\mathbf{H}_{E,P,\perp}\in\mathbb{C}^{N_E\times N_{\texttt{min}}}$ at high power is simply $\mathbf{\bar H}_E+\mathbf{T}_E\mathbf{N}_E=[\mathbf{H}_{E,P},\mathbf{H}_{E,P,\perp}]\mathbf{Q}^H$. For this reason, $\mathbf{T}_E$ and $\mathbf{H}_{E,P,\perp}$ have the same impact on DoF despite the fact that the entries of $\mathbf{T}_E$ are not i.i.d. in general. This property will be used to simplify some of the analyses shown later.

\begin{Assumption}\label{Assumption_Hij}
  $\mathbf{H}_{i,j}=\mathbf{H}_{j,i}^T$, and all entries in $\mathbf{H}_{i,j}$ for all $i<j$ over different coherence periods are i.i.d. $\mathcal{CN}(0,1)$ and independent of $\mathbf{H}_E$.
\end{Assumption}

\begin{Assumption}\label{Assumption_HE}
  All entries of $\mathbf{H}_E$ are i.i.d. $\mathcal{CN}(0,1)$.
\end{Assumption}

\begin{Assumption}\label{Assumption_Noise}
  The entries of all noise matrices at users are i.i.d. $\mathcal{CN}(0,1)$, and the entries of all noise matrices at Eve are i.i.d. $\mathcal{CN}(0,\omega^2)$.
\end{Assumption}

In fact, the value of $\omega$ has zero effect on the SDoF analysis. We can also let $\omega=1$ for convenience.

Because of Assumption \ref{Assumption_HE},  all entries in $\mathbf{H}_E'$ are also i.i.d. $\mathcal{CN}(0,1)$. In this case, all entries in the $N_E\times N_{\texttt{min}}$ matrix $\mathbf{H}_{E,P,\perp}$ remain i.i.d. $\mathcal{CN}(0,1)$ even after $\mathbf{Y}_E^{(1)}$ is available.

In phase 2, node $i$ for every $i$ sends out the $N_i\times K_2$ symbol matrix $\mathbf{X}_i$. Then node $i$ for every $i$ receives
\begin{align}\label{eq:Yi}
  &\mathbf{Y}_i^{(2)} = \sigma\sum_{j\neq i} \mathbf{H}_{i,j}\mathbf{X}_j +\mathbf{W}_i^{(2)}=\mathbf{H}_i\mathbf{X}_{(i)}+\mathbf{W}_i^{(2)}
\end{align}
with $\mathbf{X}_{(i)}$ being the vertical stack of $\mathbf{X}_j$ for $j\neq i$,
and Eve receives
\begin{align}\label{eq:YE}
  \mathbf{Y}_E^{(2)} &= \sigma\sum_{j=1}^M\mathbf{H}_{E,j} \mathbf{X}_j +\mathbf{W}_E^{(2)}=\mathbf{H}_E\mathbf{X}+\mathbf{W}_E^{(2)}
\end{align}
where $\mathbf{X}$ is the $N_T\times K_2$ matrix from the vertical stack of $\mathbf{X}_j$ for all $j$.  Let
\begin{equation}\label{eq:X'}
  \mathbf{X}'\doteq\mathbf{Q}^H\mathbf{X}=\left [\begin{array}{c}
                                              \mathbf{X}_P \\
                                              \mathbf{X}_{P,\perp}
                                            \end{array}
  \right ]
\end{equation}
 where $\mathbf{X}_P=\mathbf{Q}_P^H\mathbf{X}$ and $\mathbf{X}_{P,\perp}=\mathbf{Q}_{P,\perp}^H\mathbf{X}$. Then  \eqref{eq:YE} becomes
\begin{align}\label{eq:YEb}
  \mathbf{Y}_E^{(2)}
  &=\sigma\mathbf{H}_{E,P}\mathbf{X}_P+\sigma\mathbf{H}_{E,P,\perp}\mathbf{X}_{P,\perp}+\mathbf{W}_E^{(2)}.
\end{align}
Because of Assumption \ref{Assumption_X}, all entries in $\mathbf{X}'$ of \eqref{eq:X'} are also i.i.d. $\mathcal{CN}(0,1)$.

As a summary, we have:
\begin{Prop}\label{Property_Eve_channel}
  Under Assumptions \ref{Assumption_Pilot} and \ref{Assumption_HE} and a high SNR in phase 1, $\mathbf{Y}_i^{(1)}$ shown in \eqref{eq:Yi1} yields $\mathbf{H}_i$ uniquely, and
 $\mathbf{Y}_E^{(1)}$ shown in \eqref{eq:mYE1}, or equivalently \eqref{eq:YE1aa}, implies a unique $\mathbf{H}_{E,P}\in\mathbb{C}^{N_E\times (N_T-N_{\texttt{min}})}$ but leaves all entries of $\mathbf{H}_{E,P,\perp}\in\mathbb{C}^{N_E\times N_{\texttt{min}}}$ as i.i.d. $\mathcal{CN}(0,1)$.
\end{Prop}

\section{Decomposition of Secret-Key Capacity}\label{sec:decomposition}

Let $\mathcal{Y}_i\doteq \{\mathbf{X}_i, \mathbf{Y}_i^{(1)},\mathbf{Y}_i^{(2)}\}$ and $\mathcal{Y}_E \doteq \{\mathbf{Y}_E^{(1)},\mathbf{Y}_E^{(2)}\}$ be the data sets collected by user $i$ and Eve after the two-phase processes of multi-user ANECE. We show next that each pair-wise secret-key capacity $C_{\texttt{key},i,j}$ based on the data sets at users $i$ and $j$ from multi-user ANECE can be decomposed into two components: one for phase 1 and the other for phase 2, i.e., $C_{\texttt{key},i,j}=C_{\texttt{key},i,j}^{(1)}+C_{\texttt{key},i,j}^{(2)}$.

\subsection{Lower bound}
It follows from Lemma \ref{Lemma_key_capacity} that
\begin{equation}\label{eq:Ckeyij}
  C_{\texttt{key},i,j} \geq \mathbb{I}(\mathcal{Y}_i;\mathcal{Y}_j)-\min(C_{i,E},C_{j,E})
\end{equation}
with $C_{i,E}=\mathbb{I}(\mathcal{Y}_i;\mathcal{Y}_E)$ and
\begin{align}\label{eq:IYiYj}
  &\mathbb{I}(\mathcal{Y}_i;\mathcal{Y}_j)=\mathbb{I}(\mathbf{Y}_i^{(1)};\mathcal{Y}_j)+
  \mathbb{I}(\mathbf{X}_i,\mathbf{Y}_i^{(2)};\mathcal{Y}_j|\mathbf{Y}_i^{(1)})\notag\\
  &=\mathbb{I}(\mathbf{Y}_i^{(1)};\mathbf{Y}_j^{(1)})+
  \mathbb{I}(\mathbf{Y}_i^{(1)};\mathbf{X}_j,\mathbf{Y}_j^{(2)}|\mathbf{Y}_j^{(1)})\notag\\
  &\,\,+\mathbb{I}(\mathbf{X}_i,\mathbf{Y}_i^{(2)};\mathbf{Y}_j^{(1)}|\mathbf{Y}_i^{(1)})
  +
  \mathbb{I}(\mathbf{X}_i,\mathbf{Y}_i^{(2)};\mathbf{X}_j,\mathbf{Y}_j^{(2)}|
  \mathbf{Y}_i^{(1)},\mathbf{Y}_j^{(1)})\notag\\
  &\approx \mathbb{I}(\mathbf{Y}_i^{(1)};\mathbf{Y}_j^{(1)})+
  \mathbb{I}(\mathbf{X}_i,\mathbf{Y}_i^{(2)};\mathbf{X}_j,\mathbf{Y}_j^{(2)}|
  \mathbf{H}_i,\mathbf{H}_j).
\end{align}
The above approximation holds under a high power in phase 1, which allows $\mathbf{Y}_i^{(1)}$ to uniquely determine $\mathbf{H}_i$ for all $i$ and hence implies $\mathbb{I}(\mathbf{Y}_i^{(1)};\mathbf{X}_j,\mathbf{Y}_j^{(2)}|\mathbf{Y}_j^{(1)})\approx 0$ for all $i\neq j$. The above and all other similar approximations used in this paper do not affect the DoF analyses.
Because of $\mathbb{I}(\mathbf{Y}_i^{(1)};\mathcal{Y}_E)=0$ and $\mathbb{I}(\mathbf{X}_i,\mathbf{Y}_i^{(2)};\mathbf{Y}_E^{(1)}|\mathbf{Y}_i^{(1)})=0$ (which is due to the independence between users' channel matrices and Eve's channel matrices), we have
\begin{align}\label{eq:IYiYE}
  &C_{i,E}=\mathbb{I}(\mathbf{X}_i,\mathbf{Y}_i^{(2)};\mathcal{Y}_E|\mathbf{Y}_i^{(1)})
  =\mathbb{I}(\mathbf{X}_i,\mathbf{Y}_i^{(2)};\mathbf{Y}_E^{(2)}|\mathbf{Y}_i^{(1)},
  \mathbf{Y}_E^{(1)})
  \notag\\
  &\approx \mathbb{I}(\mathbf{X}_i,\mathbf{Y}_i^{(2)};\mathbf{Y}_E^{(2)}|\mathbf{H}_i,
  \mathbf{H}_{E,P})
\end{align}
where the approximation is due to Property \ref{Property_Eve_channel}.

We can write from \eqref{eq:Ckeyij}, \eqref{eq:IYiYj} and \eqref{eq:IYiYE} that
\begin{align}\label{eq:C_keyijlow}
  &C_{\texttt{key},i,j}\geq C_{\texttt{key},i,j}^{(1)} + \max(C_{\texttt{key},i,j,\texttt{low}}^{(2)},
  C_{\texttt{key},j,i,\texttt{low}}^{(2)})
\end{align}
where the first term is the pair-wise SKC achievable based on the data collected in phase 1 of all-user ANECE and the second term is the pair-wise SKC achievable based on the data collected in phase 2 of all-user ANECE. Here
 $C_{\texttt{key},i,j}^{(1)}=\mathbb{I}(\mathbf{Y}_i^{(1)};\mathbf{Y}_j^{(1)})$ and
\begin{equation}\label{eq:C_keyijlow_2}
  C_{\texttt{key},i,j,\texttt{low}}^{(2)}=C_{i,j}-C_{i,E},
\end{equation}
with $C_{i,j}\approx \mathbb{I}(\mathbf{X}_i,\mathbf{Y}_i^{(2)};\mathbf{X}_j,\mathbf{Y}_j^{(2)}|
  \mathbf{H}_i,\mathbf{H}_j)$.

  \subsection{Upper bound}
It also follows from Lemma \ref{Lemma_key_capacity} that with high power in phase 1,
\begin{align}\label{}
  &C_{\texttt{key},i,j}\leq \mathbb{I}(\mathcal{Y}_i;\mathcal{Y}_j|\mathcal{Y}_E)\notag\\
  &=\mathbb{I}(\mathbf{Y}_i^{(1)};\mathcal{Y}_j|\mathcal{Y}_E)+
  \mathbb{I}(\mathbf{X}_i,\mathbf{Y}_i^{(2)};\mathcal{Y}_j|\mathbf{Y}_i^{(1)},\mathcal{Y}_E)\notag\\
  &\approx
  C_{\texttt{key},i,j}^{(1)}+C_{\texttt{key},i,j,\texttt{up}}^{(2)}
\end{align}
with
\begin{equation}\label{eq:upper_bound}
  C_{\texttt{key},i,j,\texttt{up}}^{(2)}=\mathbb{I}(\mathbf{X}_i,\mathbf{Y}_i^{(2)};\mathbf{X}_j,\mathbf{Y}_j^{(2)}|
  \mathbf{Y}_i^{(1)},\mathbf{Y}_j^{(1)},\mathcal{Y}_E).
\end{equation}

Note that the phase 1 component $C_{\texttt{key},i,j}^{(1)}$ of the SKC is exact while the phase 2 component of the SKC is lower bounded by $C_{\texttt{key},i,j,\texttt{low}}^{(2)}$ and upper bounded by $C_{\texttt{key},i,j,\texttt{up}}^{(2)}$.
\subsection{Summary}
\begin{Prop}
  The pair-wise SKC in bits per channel coherence period based on the data collected from multi-user ANECE consists of two components. The first is $C_{\texttt{key},i,j}^{(1)}=\mathbb{I}(\mathbf{Y}_i^{(1)};\mathbf{Y}_j^{(1)})$, and the second is lower bounded by $C_{\texttt{key},i,j,\texttt{low}}^{(2)}$ shown in \eqref{eq:C_keyijlow_2} and upper bounded by $C_{\texttt{key},i,j,\texttt{up}}^{(2)}$ shown in \eqref{eq:upper_bound}.
\end{Prop}

We will next analyse the DoFs of $C_{\texttt{key},i,j}^{(1)}$ and $C_{\texttt{key},i,j,\texttt{low}}^{(2)}$ in two separate sections. The analysis of $C_{\texttt{key},i,j,\texttt{up}}^{(2)}$ is done in Appendix \ref{sec:upper_bound}.

\section{Phase-1 SKC of All-User ANECE}\label{sec:Phase_1}

In this section, we will derive the DoF of $C_{\texttt{key},i,j}^{(1)}$. First we can write
\begin{align}\label{eq:Csij11}
  &C_{\texttt{key},i,j}^{(1)}\doteq \mathbb{I}(\mathbf{Y}_i^{(1)};\mathbf{Y}_j^{(1)})\notag\\
  &=h(\mathbf{Y}_i^{(1)})+h(\mathbf{Y}_j^{(1)})-h(\mathbf{Y}_i^{(1)},\mathbf{Y}_j^{(1)}).
\end{align}
It follows from \eqref{eq:Yi1} that
\begin{equation}\label{eq:Yi1b}
  \mathbf{y}_i^{(1)} \doteq \texttt{vec}(\mathbf{Y}_i^{(1)})= \sigma\sum_{l\neq i} (\mathbf{P}_l^T\otimes \mathbf{I}_{N_i})\mathbf{h}_{i,l}+\mathbf{w}_i^{(1)}.
\end{equation}
Then the PDF of $\mathbf{y}_i^{(1)}$ is $f(\mathbf{y}_i^{(1)})=\mathcal{CN}(0,\mathbf{R}_{Y,i})$ with
\begin{equation}\label{eq:RYi}
  \mathbf{R}_{Y,i}
  =\left (\sum_{l\neq i} \sigma^2\mathbf{P}_l^T\mathbf{P}_l^*+\mathbf{I}_{K_1}\right )\otimes \mathbf{I}_{N_i}.
\end{equation}
To determine the joint PDF  $f(\mathbf{y}_{i,1}^{(1)},\mathbf{y}_{j,1}^{(1)})$ of $\mathbf{y}_{i,1}^{(1)}$ and $\mathbf{y}_{j,1}^{(1)}$, consider
\begin{equation}\label{eq:Yj1b'}
  {\mathbf{y}'}_j^{(1)} \doteq \texttt{vec}({\mathbf{Y}^T}_j^{(1)})= \sum_{l\neq j}\sigma \left (\mathbf{I}_{N_j}\otimes \mathbf{P}_l^T\right )\mathbf{h}_{l,j}+{\mathbf{w}'}_j^{(1)}.
\end{equation}
There is a permutation matrix $\mathbf{\Pi}_j$ such that $\mathbf{y}_j^{(1)}=\texttt{vec}(\mathbf{Y}_j^{(1)})=\mathbf{\Pi}_j{\mathbf{y}'}_j^{(1)}$.
It follows that $f(\mathbf{y}_i^{(1)},\mathbf{y}_j^{(1)})=\mathcal{CN}(0,\mathbf{R}_{Y,(i,j)})$ where $\mathbf{R}_{Y,(i,j)}$ is the covariance matrix of $[\mathbf{y}_i^{(1)T},\mathbf{y}_j^{(1)T}]^T$ which satisfies $|\mathbf{R}_{Y,(i,j)}|=|\mathbf{R}'_{Y,(i,j)}|$. Here
\begin{align}\label{}
  \mathbf{R}'_{Y,(i,j)} &\doteq\mathbb{E}\left \{ \left [\begin{array}{c}
                                                            \mathbf{y}_i^{(1)} \\
                                                            {\mathbf{y}'}_j^{(1)}
                                                          \end{array}
   \right ]\left [\begin{array}{c}
                                                            \mathbf{y}_i^{(1)} \\
                                                            {\mathbf{y}'}_j^{(1)}
                                                          \end{array}
   \right ]^H\right \}\notag\\
   &=\mathbf{\bar R}'_{Y,(i,j)}+\mathbf{I}_{(N_i+N_j)K_1},
\end{align}
with
\begin{equation}\label{eq:barRYij}
   \mathbf{\bar R}'_{Y,(i,j)}=\left [\begin{array}{cc}
                                    \mathbf{\bar R}_{Y,i}&
                                    \sigma^2\mathbf{P}_j^T\otimes \mathbf{P}_i^* \\
                                    \sigma^2\mathbf{P}_j^*\otimes \mathbf{P}_i^T &  \mathbf{\bar R}'_{Y,j}
                                  \end{array}
   \right ],
\end{equation}
and
  $\mathbf{R}'_{Y,j} =  \mathbf{I}_{N_j}\otimes(\sigma^2\sum_{l\neq j} \mathbf{P}_l^T\mathbf{P}_l^*+\mathbf{I}_{K_1})$,
$ \mathbf{\bar R}_{Y,i}
  =\left (\sigma^2\sum_{l\neq i} \mathbf{P}_l^T\mathbf{P}_l^*\right )\otimes \mathbf{I}_{N_i}$
  and $\mathbf{\bar R}'_{Y,j} =  \mathbf{I}_{N_j}\otimes(\sum_{l\neq j} \sigma^2\mathbf{P}_l^T\mathbf{P}_l^*)$.
Then, similar to an expression in the proof for Lemma \ref{Lemma:DoF_Gaussian}, we have
\begin{equation}\label{eq:Csij}
  C_{\texttt{key},i,j}^{(1)}=\log_2|\mathbf{R}_{Y,i}|+\log_2|\mathbf{R}_{Y,j}|
  -\log_2|\mathbf{R}_{Y,(i,j)}|.
\end{equation}

To study the DoF of $C_{\texttt{key},i,j}^{(1)}$, we need to remember Assumption \ref{Assumption_Pilot}.
It follows from \eqref{eq:RYi} that $\log_2|\mathbf{R}_{Y,i}|=N_i\log_2|\sum_{l\neq i} \sigma^2\mathbf{P}_l^T\mathbf{P}_l^*+\mathbf{I}_{K_1}|$. It follows from Lemma \ref{Lemma1} that with respect to $\log_2\sigma^2$,
$\texttt{DoF}(\log_2|\mathbf{R}_{Y,i}|)=N_i \texttt{rank}(\sum_{l\neq i} \sigma^2\mathbf{P}_l^T\mathbf{P}_l^*)=N_i\min(K_1,N_T-N_i)=N_i(N_T-N_i)$ where we used $K_1\geq N_T-N_{\texttt{min}}$.

To find the DoF of $\log_2|\mathbf{R}_{Y,(i,j)}|$, we need to examine the rank of $\mathbf{\bar R}'_{Y,(i,j)}$ in \eqref{eq:barRYij}, which is proportional to $\sigma^2$. We can rewrite \eqref{eq:Yi1b} as $\mathbf{y}_i^{(1)}=\sigma\mathbf{A}_i\mathbf{h}_i+\mathbf{w}_i^{(1)}$, and \eqref{eq:Yj1b'} as ${\mathbf{y}'}_j^{(1)}=\sigma\mathbf{A}'_j\mathbf{h}_j+{\mathbf{w}'}_i^{(1)}$. Clearly, both $\mathbf{A}_i$ and $\mathbf{A}_j'$ have the full column ranks $N_i(N_T-N_i)$ and $N_j(N_T-N_j)$ respectively. Note that subject to Assumption \ref{Assumption_Pilot}, every node can consistently estimate its receive channel matrices relative to other nodes. Also, there are exactly $N_iN_j$ entries in $\mathbf{h}_i$ that are identical to a corresponding set of $N_iN_j$ entries in $\mathbf{h}_j$, which are the entries in $\mathbf{H}_{i,j}=\mathbf{H}^T_{j,i}$. Other than that, all entries in $\mathbf{h}_i$ and $\mathbf{h}_j$ are statistically independent. It follows that
\begin{align}\label{}
  \mathbf{\bar R}'_{Y,(i,j)}&=\sigma^2\mathbb{E}\left \{\left [\begin{array}{c}
                                                         \mathbf{A}_i\mathbf{h}_i \\
                                                         \mathbf{A}'_j\mathbf{h}_j
                                                       \end{array}
  \right ]\left [\begin{array}{c}
                                                         \mathbf{A}_i\mathbf{h}_i \\
                                                         \mathbf{A}'_j\mathbf{h}_j
                                                       \end{array}
  \right ]^H\right \}\notag\\
  &=\sigma^2\texttt{diag}(\mathbf{A}_i,\mathbf{A}'_j)\mathbf{R}_{h,i,j}\texttt{diag}^H(\mathbf{A}_i,\mathbf{A}'_j)
\end{align}
where
\begin{equation}\label{}
  \mathbf{R}_{h,i,j}=\mathbb{E}\left \{ \left [\begin{array}{c}
                                                  \mathbf{h}_i \\
                                                  \mathbf{h}_j
                                                \end{array}
   \right ]\left [\begin{array}{c}
                                                  \mathbf{h}_i \\
                                                  \mathbf{h}_j
                                                \end{array}
   \right ]^H\right\}.
\end{equation}
For each identical pair of entries in $\mathbf{h}_i$ and $\mathbf{h}_j$, there is a corresponding row vector $\mathbf{r}_l^H$ (with two nonzero entries equal to 1 and -1 and zeros elsewhere) such that $\mathbf{r}_l^H\mathbf{R}_{h,i,j}=0$. Since there are total $N_iN_j$ such independent vectors, the rank deficiency of $\mathbf{R}_{h,i,j}$ is $N_iN_j$, i.e., $\texttt{rank}(\mathbf{R}_{h,i,j})=N_i(N_T-N_i)+N_j(N_T-N_j)-N_iN_j$. Since  $\texttt{diag}(\mathbf{A}_i,\mathbf{A}'_j)$ has the full column rank $N_i(N_T-N_i)+N_j(N_T-N_j)$,  $\texttt{rank}(\mathbf{\bar R}'_{Y,(i,j)})=\texttt{rank}(\mathbf{R}_{h,i,j})=N_i(N_T-N_i)+N_j(N_T-N_j)-N_iN_j$.

Then, it follows from \eqref{eq:Csij} that for every $i\neq j$,
\begin{align}\label{eq:DOF_Phase_1}
  &\texttt{DoF}(C_{\texttt{key},i,j}^{(1)})=\texttt{DoF}(\log_2|\mathbf{R}_{Y,i}|)\notag\\
  &\,\,+\texttt{DoF}(\log_2|\mathbf{R}_{Y,j}|)-\texttt{DoF}(\log_2|\mathbf{R}'_{Y,(i,j)}|)
  \notag\\
  &=N_iN_j.
\end{align}

Assumption \ref{Assumption_Hij} implies that the reciprocal channel matrices for different pairs of users are independent of each other for all $M\geq 2$. At high power in phase 1 of all-user ANECE, all users can obtain their reciprocal channel matrices, which can be then used to generate independent secret keys for all pairs. Hence, the pair-wise secret keys based on phase 1 of all-user ANECE are independent of each other.

As a summary, we have:
\begin{Prop}\label{Property_SDoF_Phase1}
  Under Assumptions \ref{Assumption_Pilot} and \ref{Assumption_Hij}, the SDoF between user $i$ and user $j$ from phase 1 of all-user ANECE is $N_iN_j$. Furthermore, the pair-wise secret keys from phase 1 of all-user ANECE are independent of each other at high power for all $M\geq 2$.
\end{Prop}

This property is stronger than a related result in Property 5 in \cite{WuHua2022}. The latter is based on an optimal set of pilot matrices derived in \cite{Zhu2020} subject to Assumptions \ref{Assumption_Pilot} and \ref{Assumption_Hij} and $N_i=N$ for all $i$. Property \ref{Property_SDoF_Phase1} shown above holds for all cases subject to Assumptions \ref{Assumption_Pilot} and \ref{Assumption_Hij}.

\section{Phase-2 SKC of All-User ANECE}\label{sec:Phase_2}

We now analyse $C_{\texttt{key},i,j,\texttt{low}}^{(2)}$ in \eqref{eq:C_keyijlow_2}, i.e., $C_{\texttt{key},i,j,\texttt{low}}^{(2)}=C_{i,j}-C_{i,E}$.
Note that $C_{i,j}$ defined below \eqref{eq:C_keyijlow_2} can be seen as encryption capacity between user $i$ and user $j$ via a public channel using data collected from phase 2 of all-user ANECE, and $C_{i,E}$ is the encryption leakage capacity from user $i$ to Eve via the public channel. Also note that the phase-2 SKC of ANECE does not require channel reciprocity.

\subsection{Encryption Capacity between Users}\label{sec:capacity_between_users}
It follows from the chain rule of mutual information that
\begin{align}\label{}
  &C_{i,j}\approx \mathbb{I}(\mathbf{X}_i,\mathbf{Y}_i^{(2)};\mathbf{X}_j,\mathbf{Y}_j^{(2)}|\mathbf{H}_i,\mathbf{H}_j)
  \notag\\
  &=\mathbb{I}(\mathbf{X}_i;\mathbf{X}_j,\mathbf{Y}_j^{(2)}|\mathbf{H}_i,\mathbf{H}_j)
  +\mathbb{I}(\mathbf{Y}_i^{(2)};\mathbf{X}_j,\mathbf{Y}_j^{(2)}|\mathbf{X}_i,\mathbf{H}_i,\mathbf{H}_j)\notag\\
  &=\mathbb{I}(\mathbf{X}_i;\mathbf{Y}_j^{(2)}|\mathbf{H}_i,\mathbf{H}_j)+
  \mathbb{I}(\mathbf{X}_i;\mathbf{X}_j|\mathbf{Y}_j^{(2)},\mathbf{H}_i,\mathbf{H}_j)\notag\\
  &
  +\mathbb{I}(\mathbf{Y}_i^{(2)};\mathbf{X}_j|\mathbf{X}_i,\mathbf{H}_i,\mathbf{H}_j)
  +\mathbb{I}(\mathbf{Y}_i^{(2)};\mathbf{Y}_j^{(2)}|\mathbf{X}_i,\mathbf{X}_j,\mathbf{H}_i,\mathbf{H}_j).
\end{align}
Applying the independence between $\mathbf{X}_i$ and $\mathbf{X}_j$, the independence of $\mathbf{X}_i$ for any $i$ on $\mathbf{H}_j$ for any $j$, the independence of $\mathbf{Y}_i^{(2)}$ on $\mathbf{X}_i$, and the independence of $\mathbf{Y}_j^{(2)}$ on $\mathbf{H}_i$ given $\mathbf{H}_j$, we then have
\begin{align}\label{eq:Cij}
  C_{i,j}
  &\approx \mathbb{I}(\mathbf{X}_i;\mathbf{Y}_j^{(2)}|\mathbf{H}_j)
  +\mathbb{I}(\mathbf{Y}_i^{(2)};\mathbf{X}_j|\mathbf{H}_i)\notag\\
  &
  +\mathbb{I}(\mathbf{Y}_i^{(2)};\mathbf{Y}_j^{(2)}|\mathbf{X}_i,\mathbf{X}_j,\mathbf{H}_i,\mathbf{H}_j).
\end{align}
where the third term is zero for $M=2$ but non-zero in general for $M>2$. The first two terms are similar (or symmetric). The second term can be written as
\begin{equation}\label{}
  \mathbb{I}(\mathbf{Y}_i^{(2)};\mathbf{X}_j|\mathbf{H}_i)=
  h(\mathbf{Y}_i^{(2)}|\mathbf{H}_i)-h(\mathbf{Y}_i^{(2)}|\mathbf{X}_j,\mathbf{H}_i).
\end{equation}
The 3rd term in \eqref{eq:Cij} is
\begin{align}\label{}
  &\mathbb{I}(\mathbf{Y}_i^{(2)};\mathbf{Y}_j^{(2)}|\mathbf{X}_i,\mathbf{X}_j,\mathbf{H}_i,\mathbf{H}_j)\notag\\
  &=
  h(\mathbf{Y}_i^{(2)}|\mathbf{X}_i,\mathbf{X}_j,\mathbf{H}_i,\mathbf{H}_j)-
  h(\mathbf{Y}_i^{(2)}|\mathbf{Y}_j^{(2)},\mathbf{X}_i,\mathbf{X}_j,\mathbf{H}_i,\mathbf{H}_j)\notag\\
  &=h(\mathbf{Y}_i^{(2)}|\mathbf{X}_j,\mathbf{H}_i)+
  h(\mathbf{Y}_j^{(2)}|\mathbf{X}_i,\mathbf{X}_j,\mathbf{H}_i,\mathbf{H}_j)\notag\\
  &\,\,-
  h(\mathbf{Y}_i^{(2)},\mathbf{Y}_j^{(2)}|\mathbf{X}_i,\mathbf{X}_j,\mathbf{H}_i,\mathbf{H}_j)\notag\\
  &=h(\mathbf{Y}_i^{(2)}|\mathbf{X}_j,\mathbf{H}_i)+
  h(\mathbf{Y}_j^{(2)}|\mathbf{X}_i,\mathbf{H}_j)\notag\\
  &\,\,-
  h(\mathbf{Y}_i^{(2)},\mathbf{Y}_j^{(2)}|\mathbf{X}_i,\mathbf{X}_j,\mathbf{H}_i,\mathbf{H}_j).
\end{align}
Then it is easy to verify that \eqref{eq:Cij} becomes
\begin{align}\label{eq:Cijb}
  &C_{i,j}
  \approx h(\mathbf{Y}_i^{(2)}|\mathbf{H}_i)+h(\mathbf{Y}_j^{(2)}|\mathbf{H}_j)\notag\\
  &\,\,-h(\mathbf{Y}_i^{(2)},\mathbf{Y}_j^{(2)}|\mathbf{X}_i,\mathbf{X}_j,\mathbf{H}_i,\mathbf{H}_j).
\end{align}
While Lemma \ref{Lemma:DoF_Gaussian} readily applies to the first two terms in \eqref{eq:Cijb}, the third term involves a complexity due to correlation between $\mathbf{H}_i$ and $\mathbf{H}_j$ for $M>2$.

It follows from \eqref{eq:Yi} that
\begin{equation}\label{eq:Yib}
  \mathbf{y}_i^{(2)} \doteq\texttt{vec}(\mathbf{Y}_i^{(2)}) = \sigma\sum_{j\neq i} (\mathbf{I}_{K_2}\otimes\mathbf{H}_{i,j})\mathbf{x}_j +\mathbf{w}_i.
\end{equation}
One can then verify that the PDF of $\mathbf{y}_i^{(2)}$ given $\mathbf{H}_i$ is $f(\mathbf{y}_i^{(2)}|\mathbf{H}_i)=\mathcal{CN}(0,\mathbf{R}_i)$ with
\begin{equation}\label{}
  \mathbf{R}_i\doteq\mathbb{E}\{\mathbf{y}_i^{(2)}{\mathbf{y}_i^{(2)}}^H|\mathbf{H}_i\}=
  \mathbf{I}_{K_2}\otimes (\sigma^2\mathbf{R}_{H,i}+\mathbf{I}_{N_i})
\end{equation}
with $\mathbf{R}_{H,i}= \sum_{l\neq i}\mathbf{H}_{i,l}\mathbf{H}_{i,l}^H$. Similarly,
the PDF of  $\mathbf{y}_{(i,j)}^{(2)}$ given $\mathbf{X}_i$, $\mathbf{X}_j$, $\mathbf{H}_i$ and $\mathbf{H}_j$ is $f(\mathbf{y}_{(i,j)}^{(2)}|\mathbf{X}_i,\mathbf{X}_j,\mathbf{H}_i,\mathbf{H}_j)=\mathcal{CN}(\mathbf{m}_{(i,j)},\mathbf{R}_{(i,j)})$ with ${\mathbf{y}_{(i,j)}^{(2)}}^H\doteq\left [{\mathbf{y}_i^{(2)}}^H,{\mathbf{y}_j^{(2)}}^H\right ]$, $\mathbf{m}_{(i,j)}=[\mathbf{m}_i^H,\mathbf{m}_j^H]^H$, $\mathbf{m}_i=(\mathbf{I}_{K_2}\otimes\mathbf{H}_{i,j})\mathbf{x}_j$, $\mathbf{m}_j=(\mathbf{I}_{K_2}\otimes\mathbf{H}_{j,i})\mathbf{x}_i$, and
  \begin{align}\label{}
    &\mathbf{R}_{(i,j)} \doteq\mathbb{E}\left \{(\mathbf{y}_{(i,j)}^{(2)}-\mathbf{m}_{(i,j)})(\mathbf{y}_{(i,j)}^{(2)}-\mathbf{m}_{(i,j)})^H |\mathbf{H}_i,\mathbf{H}_j\right \}\notag\\
     &=
     \left [ \begin{array}{cc}
                                   \mathbf{I}_{K_2}\otimes (\sigma^2\mathbf{R}_{H,i}+\mathbf{I}_{N_i}) & \mathbf{I}_{K_2}\otimes \sigma^2\mathbf{R}_{H,i,j} \\
                                   \mathbf{I}_{K_2}\otimes \sigma^2\mathbf{R}_{H,i,j} & \mathbf{I}_{K_2}\otimes (\sigma^2\mathbf{R}_{H,j}+\mathbf{I}_{N_j})
                                 \end{array}
    \right ]\notag\\
    &=\mathbf{\Pi}(\mathbf{I}_{K_2}\otimes (\sigma^2 \mathbf{R}_{H,(i,j)}+\mathbf{I}_{N_i+N_j})\mathbf{\Pi}^T
  \end{align}
  where $\mathbf{\Pi}$ is a permutation matrix, $\mathbf{R}_{H,i,j}=\sum_{l\notin \{i,j\}}\mathbf{H}_{i,l}\mathbf{H}_{i,l}^H$, $\mathbf{R}_{H,(i,j)}=\sum_{l\notin \{i,j\}}\mathbf{H}_{(i,j),l}\mathbf{H}_{(i,j),l}^H$, and $\mathbf{H}_{(i,j),l}=[\mathbf{H}_{i,l}^H,\mathbf{H}_{j,l}^H]^H$.
  It follows that
\begin{align}\label{eq:Cijc}
  &C_{i,j}
  \approx\mathbb{E}\{\log_2|\mathbf{R}_i|\}+\mathbb{E}\{\log_2|\mathbf{R}_j|\}
  -\mathbb{E}\{\log_2|\mathbf{R}_{(i,j)}|\}\notag\\
  &=K_2\mathbb{E}\{\log_2|\sigma^2\mathbf{R}_{H,i}+\mathbf{I}_{N_i}|\}+
  K_2\mathbb{E}\{\log_2|\sigma^2\mathbf{R}_{H,j}+\mathbf{I}_{N_j}|\}\notag\\
  &\,\,-K_2\mathbb{E}\{\log_2|\sigma^2\mathbf{R}_{H,(i,j)}+\mathbf{I}_{N_i+N_j}|\}.
\end{align}

  It follows from Assumption \ref{Assumption_Hij} that  with probability one, $\texttt{rank}(\mathbf{H}_{i,j})=\min(N_i,N_j)$, $\texttt{rank}(\sum_{l\neq i}\mathbf{H}_{i,l}\mathbf{H}_{i,l}^H)=\min(N_i,\sum_{l\neq i} N_l)$, $\texttt{rank}(\mathbf{H}_{i,j,l})=\min(N_i+N_j,N_l)$, and $\texttt{rank}(\sum_{l\notin \{i,j\}}\mathbf{H}_{(i,j),l}\mathbf{H}_{(i,j),l}^H)=\min(N_i+N_j,\sum_{l\notin \{i,j\}}N_l)$ for all $i\neq j$ and all $l\notin\{i,j\}$.
Therefore, it follows from a simple application of Corollary \ref{Corollary1} to \eqref{eq:Cijc} that the DoF of $C_{i,j}$ relative to  $\log_2\sigma^2$ is
\begin{align}\label{eq:mDoF1}
  &\texttt{DoF}(C_{i,j}) = K_2\min(N_i,N_T-N_i)+K_2\min(N_j,N_T-N_j)\notag\\
  &\,\,-K_2\min(N_i+N_j,N_T-N_i-N_j).
\end{align}

As a summary, we have:
\begin{Prop}\label{Property_Cij}
  Under Assumptions  \ref{Assumption_X}, \ref{Assumption_Hij} and \ref{Assumption_Noise}, the mutual information $C_{i,j}$ between $\{\mathbf{X}_i,\mathbf{Y}_i^{(2)}\}$ and $\{\mathbf{X}_j,\mathbf{Y}_j^{(2)}\}$ conditioned on $\{\mathbf{Y}_i^{(1)},\mathbf{Y}_j^{(1)}\}$ at high power in phase 1  is given by  \eqref{eq:Cijc} while the DoF of  $C_{i,j}$ relative to $\log_2\sigma^2$ is given by \eqref{eq:mDoF1}.
\end{Prop}

\subsection{Encryption Leakage Capacity from User $i$ to Eve}\label{sec:leakage}
Now we consider $C_{i,E}$ shown in \eqref{eq:IYiYE}, from which we have
\begin{align}\label{eq:CiE}
&C_{i,E}\approx \mathbb{I}(\mathbf{X}_i,\mathbf{Y}_i^{(2)};\mathbf{Y}_E^{(2)}|\mathbf{H}_i,
  \mathbf{H}_{E,P})\notag\\
  &=h(\mathbf{X}_i,\mathbf{Y}_i^{(2)}|\mathbf{H}_i,
  \mathbf{H}_{E,P})-h(\mathbf{X}_i,\mathbf{Y}_i^{(2)}|\mathbf{Y}_E^{(2)},\mathbf{H}_i,
  \mathbf{H}_{E,P})\notag\\
  &=h(\mathbf{X}_i|\mathbf{H}_i,
  \mathbf{H}_{E,P})+h(\mathbf{Y}_i^{(2)}|\mathbf{X}_i,\mathbf{H}_i,
  \mathbf{H}_{E,P})\notag\\
  &\,\,-h(\mathbf{X}_i,\mathbf{Y}_i^{(2)},\mathbf{Y}_E^{(2)}|\mathbf{H}_i,
  \mathbf{H}_{E,P})+h(\mathbf{Y}_E^{(2)}|\mathbf{H}_i,
  \mathbf{H}_{E,P})\notag\\
  &=h(\mathbf{Y}_i^{(2)}|\mathbf{X}_i,\mathbf{H}_i,
  \mathbf{H}_{E,P})+h(\mathbf{Y}_E^{(2)}|\mathbf{H}_i,
  \mathbf{H}_{E,P})\notag\\
  &\,\,-h(\mathbf{Y}_i^{(2)},\mathbf{Y}_E^{(2)}|\mathbf{X}_i,\mathbf{H}_i,
  \mathbf{H}_{E,P})\notag\\
  &=h(\mathbf{Y}_i^{(2)}|\mathbf{H}_i)+h(\mathbf{Y}_E^{(2)}|
  \mathbf{H}_{E,P})\notag\\
  &\,\,-h(\mathbf{Y}_i^{(2)},\mathbf{Y}_E^{(2)}|\mathbf{X}_i,\mathbf{H}_i,
  \mathbf{H}_{E,P}).
\end{align}
The first term  in \eqref{eq:CiE} is the same as that in \eqref{eq:Cijb}, for which we have a simple expression due to the Gaussian distribution of $\mathbf{Y}_i^{(2)}$ given $\mathbf{H}_i$, which (by Lemma \ref{Lemma:DoF_Gaussian}) leads to
\begin{equation}\label{eq:DoFhYiHi}
  \texttt{DoF}(h(\mathbf{Y}_i^{(2)}|\mathbf{H}_i))=\min(N_i,N_T-N_i)K_2.
\end{equation}
But the other two terms in \eqref{eq:CiE} are more involved because $\mathbf{Y}_E^{(2)}$ is not Gaussian distributed while conditioned on $\mathbf{H}_{E,P}$. Notice the second term in \eqref{eq:YEb}, which is a product of two Gaussian distributed matrices. While finding the simpler forms of $h(\mathbf{Y}_E^{(2)}|
  \mathbf{H}_{E,P})$ and $h(\mathbf{Y}_i^{(2)},\mathbf{Y}_E^{(2)}|\mathbf{X}_i,\mathbf{H}_i,
  \mathbf{H}_{E,P})$ remains an open challenge, we focus next on the DoF of these two terms.

 \subsubsection{Analysis of $h(\mathbf{Y}_E^{(2)}|\mathbf{H}_{E,P})$}\label{sec:hYEHEP}
Recall \eqref{eq:YEb} and define
\begin{equation}\label{eq:alpha_beta}
  \mathbf{Y}_E^{(2)}=[
\mathbf{Y}_{E,\alpha}^{(2)},
 \mathbf{Y}_{E,\beta}^{(2)}
 ] =\left [ \begin{array}{cc}
                               \mathbf{Y}_{E,\alpha,a}^{(2)},\mathbf{Y}_{E,\beta,a}^{(2)} \\
                               \mathbf{Y}_{E,\alpha,a}^{(2)},\mathbf{Y}_{E,\beta,b}^{(2)}
                             \end{array}
  \right ]
\end{equation}
where $\mathbf{Y}_{E,\alpha}^{(2)}$ has $\min(N_{\texttt{min}},K_2)$ columns, $\mathbf{Y}_{E,\beta}^{(2)}$ has $\Delta K_2=(K_2-N_{\texttt{min}})^+$ columns, $[\mathbf{Y}_{E,\alpha,a}^{(2)},\mathbf{Y}_{E,\beta,a}^{(2)}]$ has $\min(N_T,N_E)$ rows, and $[\mathbf{Y}_{E,\alpha,b}^{(2)},\mathbf{Y}_{E,\beta,b}^{(2)}]$ has $\Delta N_E=(N_E-N_T)^+$ rows.

It follows that
\begin{align}\label{eq:new1}
&\texttt{DoF}(h(\mathbf{Y}_{E,\alpha}^{(2)}|\mathbf{H}_{E,P}))\geq
\texttt{DoF}(h(\mathbf{Y}_{E,\alpha}^{(2)}|\mathbf{H}_{E,P},\mathbf{X}))\notag\\
&=N_E\min(N_{\texttt{min}},K_2)
\end{align}
where the last equality is due to the unknown $\mathbf{H}_{E,P,\perp}$. Also
the equality in the above ``$\geq$'' holds since $\mathbf{Y}_{E,\alpha}^{(2)}$ has total $N_E\min(N_{\texttt{min}},K_2)$ elements.

If $K_2>N_{\texttt{min}}$, we can write
\begin{equation}\label{}
  \mathbf{Y}_{E,T}\doteq[\mathbf{Y}_E^{(1)},\mathbf{Y}_{E,\alpha}^{(2)}]
  =\sigma\mathbf{H}_E\mathbf{P}_T+\mathbf{W}_{E,T}
\end{equation}
with $\mathbf{P}_T=[\mathbf{P},\mathbf{X}_\alpha]\in\mathbb{C}^{N_T\times N_T}$ and $\mathbf{X}_\alpha$ being the first $N_{\texttt{min}}$ columns of $\mathbf{X}=[\mathbf{X}_\alpha,\mathbf{X}_\beta]$. At high power, $\mathbf{H}_E\approx \frac{1}{\sigma}\mathbf{Y}_{E,T}\mathbf{P}_T^{-1}$. It follows that
\begin{align}\label{}
  &\mathbf{Y}_{E,\beta,a}^{(2)}\approx\sigma\mathbf{H}_{E,a}\mathbf{X}_\beta
  =\mathbf{Y}_{E,T,a}\mathbf{P}_T^{-1}\mathbf{X}_\beta
\end{align}
where we have used the partitions  $\mathbf{Y}_{E,T}=[\mathbf{Y}_{E,T,a}^T,\mathbf{Y}_{E,T,b}^T]^T$ and $\mathbf{H}_E=[\mathbf{H}_{E,a}^T,\mathbf{H}_{E,b}^T]^T$, which are compatible with $\mathbf{Y}_{E,\alpha}=[\mathbf{Y}_{E,\alpha,a}^T,\mathbf{Y}_{E,\alpha,b}^T]^T$.
Furthermore,
\begin{align}\label{eq:new2}
&\texttt{DoF}(h(\mathbf{Y}_{E,\beta,a}^{(2)}|\mathbf{Y}_{E,\alpha}^{(2)},\mathbf{H}_{E,P}))
=\texttt{DoF}(h(\mathbf{Y}_{E,\beta,a}^{(2)}|\mathbf{Y}_{E,\alpha}^{(2)},\mathbf{Y}_E^{(1)}))\notag\\
&=\texttt{DoF}(h(\mathbf{Y}_{E,\beta,a}^{(2)}|\mathbf{Y}_{E,T}))\geq \texttt{DoF}(h(\mathbf{Y}_{E,\beta,a}^{(2)}|\mathbf{Y}_{E,T},\mathbf{X}_\alpha))\notag\\
&=\texttt{DoF}(h(\mathbf{Y}_{E,\beta,a}^{(2)}|\mathbf{H}_E))
=\min(N_E,N_T)\Delta K_2
\end{align}
where the last equality is due to the unknown $\mathbf{X}_\beta$. Also
the equality in the above ``$\geq$'' holds since $\mathbf{Y}_{E,\beta,a}^{(2)}$ has $\min(N_E,N_T)\Delta K_2$ elements.

If $K_2>N_{\texttt{min}}$ and $N_E>N_T$, we can write $\mathbf{Y}_{E,\beta,b}^{(2)}\approx
\mathbf{Y}_{E,T,b}\mathbf{P}_T^{-1}\mathbf{X}_\beta\approx
\mathbf{Y}_{E,T,b}\mathbf{Y}_{E,T,a}^{-1}\mathbf{Y}_{E,\beta,a}^{(2)}
$ and hence
\begin{equation}\label{eq:new3}
  \texttt{DoF}(h(\mathbf{Y}_{E,\beta,b}^{(2)}|\mathbf{Y}_{E,\beta,a}^{(2)},
  \mathbf{Y}_{E,\alpha}^{(2)},\mathbf{H}_{E,P}))=0.
\end{equation}

Adding up \eqref{eq:new1}, \eqref{eq:new2} and \eqref{eq:new3} yields
\begin{align}\label{eq:DoFhYEHEP}
&\texttt{DoF}(h(\mathbf{Y}_E^{(2)}|\mathbf{H}_{E,P}))\notag\\
&=N_E\min(N_{\texttt{min}},K_2)+\min(N_E,N_T)\Delta K_2.
\end{align}

\subsubsection{Analysis of $h(\mathbf{Y}_i^{(2)},\mathbf{Y}_E^{(2)}|\mathbf{X}_i,
  \mathbf{H}_i,\mathbf{H}_{E,P})$}\label{sec:hYiYEXiHiHEP}
  For convenience, let us consider $h(\mathbf{Y}_1^{(2)},\mathbf{Y}_E^{(2)}|\mathcal{C}_1)$ without loss of generality. Here $\mathcal{C}_1 = \{\mathbf{X}_1,
  \mathbf{H}_{1,P},\mathbf{H}_{1,P,\perp},\mathbf{H}_{E,P}\}$.
  It follows that
  \begin{align}\label{eq:new1b}
  &\texttt{DoF}(h(\mathbf{Y}_1^{(2)}|\mathcal{C}_1))=\min(N_1,N_{T,1})K_2
  \end{align}
  with $N_{T,1}=N_T-N_1$. Given high power, $\mathcal{C}_1$ and $\mathbf{Y}_1^{(2)}\approx \sigma \mathbf{H}_1\mathbf{X}_{(1)}$, we have $\mathbf{X}_{(1)}\approx
  \mathbf{X}_{1,0}+\mathbf{N}_1\mathbf{T}_1$, where $\mathbf{X}_{1,0}$ and $\mathbf{N}_1$ are given, the column span of $\mathbf{N}_1\in \mathbb{C}^{N_{T,1}\times (N_{T,1}-N_1)^+}$ is the right null space of $\mathbf{H}_1\in\mathbb{C}^{N_1\times N_{T,1}}$, and $\mathbf{T}_1\in\mathbb{C}^{(N_{T,1}-N_1)^+\times K_2}$ is arbitrary.  So,
  \begin{align}
  &\mathbf{Y}_E^{(2)} \approx \sigma \mathbf{H}_E\mathbf{X}\approx \sigma\mathbf{H}_E\left [ \begin{array}{c}
                                              \mathbf{X}_1 \\
                                              \mathbf{X}_{1,0}+\mathbf{N}_1\mathbf{T}_1
                                            \end{array}
  \right ].
  \end{align}

  Recall the $(\alpha,\beta)$-partitions $\mathbf{Y}_E^{(2)}=[\mathbf{Y}_{E,\alpha}^{(2)},\mathbf{Y}_{E,\beta}^{(2)}]$ with
   $\mathbf{Y}_{E,\alpha}^{(2)}\in\mathbb{C}^{N_E\times \min(N_{\texttt{min}},K_2)}$ and $\mathbf{Y}_{E,\beta}^{(2)}
   \in\mathbb{C}^{N_E\times \Delta K_2}$ and $\Delta K_2=(K_2-N_{\texttt{min}})^+$. Similar to \eqref{eq:new1},
   \begin{equation}\label{eq:new2b}
     \texttt{DoF}(h(\mathbf{Y}_{E,\alpha}^{(2)}|\mathbf{Y}_1^{(2)},\mathcal{C}_1))=
  N_E\min(N_{\texttt{min}},K_2).
   \end{equation}

 For $K_2>N_{\texttt{min}}$, recall $\mathbf{Y}_{E,T} = [\mathbf{Y}_E^{(1)},\mathbf{Y}_{E,\alpha}^{(2)}]\approx\mathbf{H}_E\mathbf{P}_T$ and $\mathbf{H}_E\approx \mathbf{Y}_{E,T}\mathbf{P}_T^{-1}$. Note that $\mathbf{P}_T$ is dependent on $\mathbf{X}_\alpha$ in $\mathbf{X}=[\mathbf{X}_\alpha,\mathbf{X}_\beta]$. More specifically, given $\{\mathbf{Y}_{E,\alpha}^{(2)},\mathbf{Y}_1^{(2)},\mathcal{C}_1\}$ and high power, $\mathbf{H}_E$ is a (uniquely valued) function of $\mathbf{T}_{1,\alpha}\in\mathbb{C}^{(N_{T,1}-N_1)^+\times N_{\texttt{min}}}$ in $\mathbf{T}_1=[\mathbf{T}_{1,\alpha},\mathbf{T}_{1,\beta}]$. We see that if $(N_{T,1}-N_1)^+=0$, there is no more freedom in $\mathbf{Y}_{E,\beta}^{(2)}$ since both $\mathbf{H}_E$ and $\mathbf{X}$ are given at high power.

 If $(N_{T,1}-N_1)^+>0$, we now define the $(a,b)$-partitions of $\mathbf{Y}_{E,\beta}^{(2)}$ as $\mathbf{Y}_{E,\beta}^{(2)}=[\mathbf{Y}_{E,\beta,a}^{(2)T},\mathbf{Y}_{E,\beta,b}^{(2)T}]^T$ with $\mathbf{Y}_{E,\beta,a}^{(2)}\in\mathbb{C}^{\min(N_E,(N_{T,1}-N_1)^+)\times \Delta K_2}$ and $\mathbf{Y}_{E,\beta,b}^{(2)}\in\mathbb{C}^{(N_E-(N_{T,1}-N_1)^+)^+\times \Delta K_2}$. Then at high power, we can write
\begin{equation}\label{eq:CBT}
  \mathbf{Y}_{E,\beta,a}^{(2)}\approx\mathbf{C}_{\beta,a}+\mathbf{B}_a\mathbf{T}_{1,\beta}
\end{equation}
where $\mathbf{C}_{\beta,a}$ and $\mathbf{B}_a$ are a function of $\mathbf{T}_{1,\alpha}$ but independent of $\mathbf{T}_{1,\beta}$. Note that $\mathbf{T}_1=[\mathbf{T}_{1,\alpha},\mathbf{T}_{1,\beta}]$. It follows that
\begin{align}\label{eq:new3b}
&\texttt{DoF}(h(\mathbf{Y}_{E,\beta,a}^{(2)}|\mathbf{Y}_{E,\alpha}^{(2)},\mathbf{Y}_1^{(2)},
\mathcal{C}_1))\notag\\
&\geq \texttt{DoF}(h(\mathbf{Y}_{E,\beta,a}^{(2)}|\mathbf{T}_{1,\alpha},
\mathbf{Y}_{E,\alpha}^{(2)},\mathbf{Y}_1^{(2)},
\mathcal{C}_1))\notag\\
&=\min(N_E,(N_{T,1}-N_1)^+)\Delta K_2
\end{align}
where the last equality is due to the unknown $\mathbf{T}_{1,\beta}$. Also the equality in the above ``$\geq$'' holds because of the fullness of DoF (i.e., it equals the number of entries in $\mathbf{Y}_{E,\beta,a}^{(2)}$).

If $K_2>N_{\texttt{min}}$ and $N_E>(N_{T,1}-N_1)^+>0$, given $\{\mathbf{Y}_{E,\beta,a}^{(2)},
\mathbf{Y}_{E,\alpha}^{(2)},\mathbf{Y}_1^{(2)},
\mathcal{C}_1\}$ implies that $\mathbf{H}_E$ is a function of $\mathbf{T}_{1,\alpha}$ but now conditioned by another random realization of $\mathbf{T}_{1,\beta}$. This makes $\mathbf{H}_E$ deterministic with probability one at high power. A constant $\mathbf{H}_E$ (or equivalently constant $\mathbf{C}_{\beta,a}$ and $\mathbf{B}_a\in\mathbb{C}^{(N_{T,1}-N_1)\times (N_{T,1}-N_1)}$ in \eqref{eq:CBT}) combined with the knowledge of $\mathbf{Y}_{E,\beta,a}^{(2)}$ also makes $\mathbf{T}_{1,\beta}$ constant at high power. This in turn makes $\mathbf{Y}_{E,\beta,b}^{(2)}$ constant. Therefore,
\begin{align}\label{eq:new4b}
&\texttt{DoF}(h(\mathbf{Y}_{E,\beta,b}^{(2)}|\mathbf{Y}_{E,\beta,a}^{(2)},
\mathbf{Y}_{E,\alpha}^{(2)},\mathbf{Y}_1^{(2)},
\mathcal{C}_1))=0.
\end{align}

  Adding up \eqref{eq:new1b}, \eqref{eq:new2b}, \eqref{eq:new3b} and \eqref{eq:new4b} but with $N_1$ replaced by $N_i$ yields
  \begin{align}\label{eq:DoFhYiYEXiHiHEP}
 & \texttt{DoF}(h(\mathbf{Y}_i^{(2)},\mathbf{Y}_E^{(2)}|\mathbf{X}_i,
  \mathbf{H}_i,\mathbf{H}_{E,P}))\notag\\
  &=K_2\min(N_i,N_T-N_i)+N_E\min(N_{\texttt{min}},K_2)\notag\\
  &\,\,+\Delta K_2\min(N_E,(N_T-2N_i)^+).
  \end{align}

  \subsubsection{Summary} Combining the above results, we have:
  \begin{Prop}\label{Property_DoF_leakage}
    The DoF of the encryption leakage capacity $C_{i,E}$ is given by \texttt{Eq.}\eqref{eq:DoFhYiHi}+\texttt{Eq.}\eqref{eq:DoFhYEHEP}-\texttt{Eq.}\eqref{eq:DoFhYiYEXiHiHEP}.
  \end{Prop}
\subsection{Phase-2 SDoF of All-User ANECE}
It follows from Properties \ref{Property_Cij} and \ref{Property_DoF_leakage} (i.e., \eqref{eq:mDoF1} \eqref{eq:CiE}, \eqref{eq:DoFhYiHi}, \eqref{eq:DoFhYEHEP} and \eqref{eq:DoFhYiYEXiHiHEP})  that  the DoF of $C_{\texttt{key},i,j,\texttt{low}}^{(2)}=C_{i,j}-C_{i,E}$ is
\begin{align}\label{eq:SDoF_low}
&\texttt{DoF}(C_{\texttt{key},i,j,\texttt{low}}^{(2)}) = K_2\min(N_j,N_T-N_j)\notag\\
&\,\,+K_2\min(N_i,N_T-N_i)+\Delta K_2\min(N_E,(N_T-2N_i)^+)\notag\\
&\,\,-K_2\min(N_i+N_j,N_T-N_i-N_j)\notag\\
&\,\,-\Delta K_2\min(N_E,N_T)
\end{align}
with  $\Delta K_2=(K_2-N_{\texttt{min}})^+$. The above is a non-increasing function of $N_E$. Since $\texttt{DoF}(C_{\texttt{key},i,j,\texttt{low}}^{(2)})$ could be negative, we will also use $\texttt{DoF}^+(C_{\texttt{key},i,j,\texttt{low}}^{(2)})=
\max(0,\texttt{DoF}(C_{\texttt{key},i,j,\texttt{low}}^{(2)}))$ when it is needed.

It follows from the results shown in Appendix \ref{sec:upper_bound} (i.e., \eqref{eq:upper_bound4}, \eqref{eq:DoFhYEHEP}, \eqref{eq:DoFhYiYEXiHiHEP} and \eqref{eq:DoFhYiYjYEXiXj}) that
\begin{align}\label{eq:SDoF_upB}
& \texttt{DoF}(C_{\texttt{key},i,j,\texttt{up}}^{(2)})\notag\\
&=K_2\min(N_i,N_T-N_i)+K_2\min(N_j,N_T-N_j)\notag\\
&\,\,+\Delta K_2\min(N_E,(N_T-2N_i)^+)\notag\\
&\,\,+\Delta K_2\min(N_E,(N_T-2N_j)^+)\notag\\
&\,\,-\Delta K_2\min(N_E,N_T)\notag\\
&\,\,-\Delta K_2\min(N_E,(N_T-2N_i-2N_j)^+)\notag\\
&\,\,-K_2\min(N_i,N_T-N_i-N_j)\notag\\
&\,\,-K_2\min(N_j,(N_T-2N_i-N_j)^+).
\end{align}

Furthermore, one can verify that the gap between the upper and lower bounds is
\begin{align}\label{eq:SDoF_gap}
&G_{i,j}\doteq \texttt{DoF}(C_{\texttt{key},i,j,\texttt{up}}^{(2)})-
\texttt{DoF}(C_{\texttt{key},i,j,\texttt{low}}^{(2)})\notag\\
&=\Delta K_2\min(N_E,(N_T-2N_j)^+)\notag\\
&\,\,+K_2\min(N_i+N_j,N_T-N_i-N_j)\notag\\
&\,\,-K_2\min(N_i,N_T-N_i-N_j)\notag\\
&\,\,-K_2\min(N_j,(N_T-2N_i-N_j)^+)\notag\\
&\,\,-\Delta K_2\min(N_E,(N_T-2N_i-2N_j)^+).
\end{align}
Note that the upper bound is always symmetric, i.e.,
 $C_{\texttt{key},i,j,\texttt{up}}^{(2)}=C_{\texttt{key},j,i,up}^{(2)}$.

\subsubsection{For a symmetric multi-user network}
We now consider a multi-user symmetric network where $N_i=N$.
One can verify that \eqref{eq:SDoF_gap} reduces to
\begin{equation}\label{}
  G_{i,j}=\left \{ \begin{array}{cc}
                     0, & M=2; \\
                     \Delta K_2\min(N_E,N), & M=3; \\
                     \Delta K_2\min(N_E,2N),&M= 4;\\
                     0,&M\geq \lceil 4+\frac{N_E}{N}\rceil.
                   \end{array}
  \right .
\end{equation}
with $\Delta K_2=K_2-\min(N,K_2)=(K_2-N)^+$.

Furthermore, for $M\geq 4$,
the gap is
\begin{align}\label{}
  &G_{i,j}=\Delta K_2 [\min(N_E,(M-2)N)\notag\\
  &\,\,-\min(N_E,(M-4)N)]\geq 0
\end{align}
with equality if $M\geq \lceil4+\frac{N_E}{N}\rceil$.
Also, under $M\geq \lceil4+\frac{N_E}{N}\rceil$, one can verify from \eqref{eq:SDoF_low} and \eqref{eq:SDoF_upB} that
\begin{align}\label{eq:SDoF_large_M}
  &\texttt{DoF}(C_{\texttt{key},i,j,\texttt{up}}^{(2)})
  =\texttt{DoF}(C_{\texttt{key},i,j,\texttt{low}}^{(2)})=0.
\end{align}

One can also verify that for $N_E\geq MN$, \eqref{eq:SDoF_low} reduces to
\begin{align}\label{eq:69}
  &\texttt{DoF}^+(C_{\texttt{key},i,j,\texttt{low}}^{(2)})\notag\\
  &=\left \{ \begin{array}{cc}
               2N\min(N,K_2), & M=2; \\
               N(2\min(N,K_2)-K_2)^+, & M=3; \\
               (-2\Delta K_2N)^+=0, & M\geq 4;
             \end{array}
  \right .
\end{align}
and \eqref{eq:SDoF_upB} reduces to $\texttt{DoF}(C_{\texttt{key},i,j,\texttt{up}}^{(2)})=0$ for $M\geq 4$. The above \eqref{eq:69} reaches its maximum when $K_2=N$.
We also see that if $N_E\geq MN$, the above \eqref{eq:69} is positive for ``$M=2$ and $K_2\geq 1$'' and for ``$M=3$ and $1\leq K_2<2N$''.

It also follows that if $K_2=N$, then for all $N_E\geq 0$,
\begin{equation*}
  \texttt{DoF}(C_{\texttt{key},i,j,\texttt{low}}^{(2)})=
\texttt{DoF}(C_{\texttt{key},i,j,\texttt{up}}^{(2)})=\left \{\begin{array}{cc}
                                                               2N^2, & M=2; \\
                                                               N^2, & M=3.
                                                             \end{array}
 \right .
\end{equation*}

\subsubsection{For asymmetric two-user network}
For $M=2$ and $N_1\leq N_2$, one can verify that $G_{1,2}=0$ and $G_{2,1}=\Delta K_2\min(N_E,N_2-N_1)$. This means that a public transmission from use 1 to user 2 using the data from phase 2 of two-user ANECE achieves the optimal SDoF. Furthermore, in this case, \eqref{eq:SDoF_upB} reduces to
\begin{align}\label{eq:SDoF_upB2}
& \texttt{SDoF}_{\texttt{original}}^{(2)}\doteq \texttt{DoF}(C_{s,1,2,\texttt{up}}^{(2)})=\texttt{DoF}(C_{s,1,2,\texttt{low}}^{(2)})\notag\\
&=\left \{\begin{array}{cc}
            2K_2N_1, & \texttt{C1}; \\
            2K_2N_1-\Delta K_2(N_E-\Delta N), & \texttt{C2}; \\
            2\min(N_1,K_2)N_1, & \texttt{C3};
          \end{array}
 \right .
\end{align}
with $\Delta N=N_2-N_1$ and $\Delta K_2=(K_2-N_1)^+$. And ``\texttt{C1}'', ``\texttt{C2}'' and ``\texttt{C3}'' are  respectively ``$0\leq N_E\leq \Delta N$'', ``$\Delta N\leq N_E\leq N_T=N_1+N_2$'' and ``$N_E\geq N_T$'', which are ``small, medium and large'' regions of $N_E$.
\subsection{Summary}
\begin{Prop}\label{Property_major} The phase-2 SDoF of multi-user ANECE has the following properties:

 1) The phase-2 SDoF of $M$-user ANECE is lower bounded by \eqref{eq:SDoF_low} and upper bounded by \eqref{eq:SDoF_upB}.

2) For a symmetric network with  $N_i=N$ for all $i=1,\cdots,M$, the gap between the upper and lower bounds is given by \eqref{eq:SDoF_gap}, which is zero if $K_2\leq N$, or if
 $M=2$, or if
     $M\geq 4+\lceil \frac{N_E}{N}\rceil$.

3)  For $M=2$ and $N_1\leq N_2$, the phase-2 SDoF is given by \eqref{eq:SDoF_upB2}, which equals $2N_1^2$ if $N_E\geq N_T$ and $K_2\geq N_1$.

4) For a symmetric network of $M\geq 3$ users, the phase-2 SDoF  is zero if $M\geq 4+\lceil \frac{N_E}{N}\rceil$, or if $N_E\geq MN$ and $M\geq 4$. But for $M=3$, the phase-2 SDoF is positive subject to $1\leq K_2<2N$ even if $N_E\geq 3N$. Furthermore, if using $M=3$ and $K_2=N$, the phase-2 SDoF equals $N^2$ for all $N_E\geq 0$.
\end{Prop}

\section{Pair-Wise ANECE for Multi-User Network}\label{sec:Pairwise}
We now consider the use of pair-wise ANECE in a network of $M\geq 3$ nodes (or users) with $N_i$ antennas on the $i$th node. We will show that the pair-wise ANECE has a serious disadvantage when compared with all-user ANECE.

Assume that within each coherence period, all pairs of users sequentially conduct the two-phase sessions of 2-user (i.e., pair-wise) ANECE. The signal received by user $i_p$ in phase 1 of the $p$th session is then
\begin{equation}\label{}
  \mathbf{Y}_{i_p}^{(1)}=\sigma\mathbf{H}_{i_p,j_p}\mathbf{P}_{j_p}+\mathbf{W}_{i_p}^{(1)}
\end{equation}
where $p=1,2,\cdots,P_0$, $P_0=\frac{1}{2}M(M-1)$, $i_p\neq j_p$, and $(i_p,j_p)$ denotes the pair of node $i_p$ and node $j_p$. Assume that $\mathbf{P}_{j_p}\in \mathbb{C}^{N_{j_p}\times k_1}$ has the full row rank for all $j_p\in \{1,\cdots,M\}$. Then node $i_p$ can obtain a consistent estimate of $\mathbf{H}_{i_p,j_p}$ from $\mathbf{Y}_{i_p}^{(1)}$. Here $P_0$ is the number of pairs among $M$ users.

The signals received by Eve in phase 1 of all $P_0$ sessions can be now grouped into (up to a known permutation):
\begin{equation}\label{}
  \mathbf{Y}_{E,pair}^{(1)}=\sigma\mathbf{H}_E\mathbf{P}_{\texttt{pair}}+\mathbf{W}_E^{(1)}
\end{equation}
where $\mathbf{Y}_E^{(1)}=[\mathbf{Y}_{E,1}^{(1)},\cdots,\mathbf{Y}_{E,P_0}^{(1)}]$, $\mathbf{H}_E=[\mathbf{H}_{E,1}, \cdots, \mathbf{H}_{E,M}]$, $\mathbf{W}_E^{(1)}=[\mathbf{W}_{E,1}^{(1)},\cdots,\mathbf{W}_{E,P_0}^{(1)}]$, and
\begin{align}\label{}
\mathbf{P}_{\texttt{pair}}&=\left [ \begin{array}{cccc}
                                      \mathbf{P}_1 & \mathbf{P}_1 & 0 & \cdots\\
                                      \mathbf{P}_2 &0 & \mathbf{P}_2 & \cdots \\
                                      0 & \mathbf{P}_3 & \mathbf{P}_3 & \cdots \\
                                      \cdots & \cdots & \cdots & \cdots
                                    \end{array}
  \right ].
\end{align}
The matrix $\mathbf{P}_{\texttt{pair}}$ has $M\times P_0$ blocks as illustrated above and has the full-row rank $N_T$ provided $M\geq 3$. Hence Eve can also obtain a consistent estimate of $\mathbf{H}_E$ in this case.

\subsection{Analysis of Phase 1}
Since phase 1 of each session requires $k_1$ sampling intervals (or time slots), the total sampling intervals required for all sessions within a coherence period is $P_0k_1$. Provided that the full row-rank condition of $\mathbf{P}_{j_p}$ is met for all $p$, every pair of users can obtain a consistent estimate of their reciprocal channel matrix, and hence the DoF of the secret-key capacity from phase 1 of session $p$ is given by $N_{i_p}N_{j_p}$, which is the same as \eqref{eq:DOF_Phase_1}.

For a further comparison, let $N_i=N$ for all $i$. Then the pair-wise ANECE requires a total of $P_0N=\frac{1}{2}M(M-1)N$ time slots for phase-1 transmissions. But the multi-user ANECE requires only $(M-1)N$ time slots for phase-1 transmission.

\subsection{Analysis of Phase 2}
In phase 2 of the $p$th session,  users $i_p$ and $j_p$ transmit $\mathbf{X}_{i_p}\in\mathbb{C}^{N_{i_p}\times k_2}$ and $\mathbf{X}_{j_p}\in\mathbb{C}^{N_{j_p}\times k_2}$ respectively, and they also receive respectively
\begin{equation}\label{}
  \mathbf{Y}_{i_p}^{(2)}=\sigma\mathbf{H}_{i_p,j_p}\mathbf{X}_{j_p}+\mathbf{W}_{i_p}^{(2)}
\end{equation}
and $\mathbf{Y}_{j_p}^{(2)}=\sigma\mathbf{H}_{i_p,j_p}^T\mathbf{X}_{i_p}+\mathbf{W}_{j_p}^{(2)}$. At the same time,
 Eve receives
\begin{equation}\label{eq:YEp2}
  \mathbf{Y}_{E,p}^{(2)} = \sigma[\mathbf{H}_{E,i_p},\mathbf{H}_{E,j_p}]\left [\begin{array}{c}
                                                                       \mathbf{X}_{i_p} \\
                                                                       \mathbf{X}_{j_p}
                                                                     \end{array}
   \right ]+\mathbf{W}_{E,p}^{(2)}
\end{equation}
The secret-key capacity $C_{key,i_p,j_p}^{(2)}$ between user $i_p$ and user $j_p$ using measurements in phase 2  is lower bounded by
\begin{equation}\label{}
  C_{s,i_p,j_p,\texttt{low}}^{(2)}=C_{i_p,j_p}-C_{i_p,E}
\end{equation}
 where
 \begin{equation}\label{}
   C_{i_p,j_p}\approx \mathbb{I}(\mathbf{X}_{i_p};\mathbf{Y}_{j_p}^{(2)}|\mathbf{H}_{j_p,i_p})
  +\mathbb{I}(\mathbf{X}_{j_p};\mathbf{Y}_{i_p}^{(2)}|\mathbf{H}_{i_p,j_p})
 \end{equation}
 which is similar to \eqref{eq:Cij} where the last term is zero due to $M=2$. It follows that $\texttt{DoF}(\mathbb{I}(\mathbf{X}_{i_p};\mathbf{Y}_{j_p}^{(2)}|\mathbf{H}_{j_p,i_p}))
 =\texttt{DoF}(h(\mathbf{Y}_{j_p}^{(2)}|\mathbf{H}_{j_p,i_p}))=\min(N_{i_p},N_{j_p})k_2$. The second term is symmetric to the first term.
  Also
 \begin{align}\label{eq:CiE_pair}
&C_{i_p,E}\approx h(\mathbf{Y}_{i_p}^{(2)}|\mathbf{H}_{i_p,j_p})+h(\mathbf{Y}_{E,p}^{(2)}|
  \mathbf{H}_E)\notag\\
  &\,\,-h(\mathbf{Y}_{i_p}^{(2)},\mathbf{Y}_{E,p}^{(2)}|\mathbf{X}_{i_p},\mathbf{H}_{i_p,j_p},
  \mathbf{H}_E).
\end{align}
which is similar to \eqref{eq:CiE} but here the entire $\mathbf{H}_E$ is known to Eve. We know that $\texttt{DoF}(h(\mathbf{Y}_{i_p}^{(2)}|\mathbf{H}_{i_p,j_p}))=\min(N_{i_p},N_{j_p})k_2$ and $h(\mathbf{Y}_{E,p}^{(2)}|
  \mathbf{H}_E)=\min(N_E,N_{i_p}+N_{j_p})k_2$. For the third term in \eqref{eq:CiE_pair}, we can write
 \begin{align}
&\left [\begin{array}{c}
          \mathbf{Y}_{i_p}^{(2)} \\
          \mathbf{Y}_{E,p}^{(2)}
        \end{array}
 \right ]=\sigma\left [ \begin{array}{c}
                    0 \\
                    \mathbf{H}_{E,i_p}
                  \end{array}
 \right ]\mathbf{X}_{i_p}+\sigma\left [ \begin{array}{c}
                                    \mathbf{H}_{i_p,j_p} \\
                                    \mathbf{H}_{E,j_p}
                                  \end{array}
 \right ]\mathbf{X}_{j_p}\notag\\
 &\,\,+\left [\begin{array}{c}
                                   \mathbf{W}_{i_p}^{(2)} \\
                                   \mathbf{W}_{E,p}^{(2)}
                                 \end{array}
 \right ]
\end{align}
where $rank\left [ \begin{array}{c}
                                    \mathbf{H}_{i_p,j_p} \\
                                    \mathbf{H}_{E,j_p}
                                  \end{array}
 \right ]=\min(N_E+N_{i_p},N_{j_p})$.
It follows that $\texttt{DoF}(h(\mathbf{Y}_{i_p}^{(2)},\mathbf{Y}_{E,p}^{(2)}|\mathbf{X}_{i_p},
  \mathbf{H}_{i_p,j_p},\mathbf{H}_E))=\min(N_E+N_{i_p},N_{j_p})k_2$.

Combing the above results, we have
\begin{align}\label{eq:SDoF_pair_xx}
  &\texttt{DoF}(C_{s,i_p,j_p,\texttt{low}}^{(2)}) = \min(N_{i_p},N_{j_p})k_2\notag\\
  &\,\,-\min(N_E,N_{i_p}+N_{j_p})k_2+\min(N_E+N_{i_p},N_{j_p})k_2
\end{align}
where $k_2$ is the number of time slots used for phase 2 of each session.  For comparison with all-user ANECE, we will choose $k_2=\frac{1}{P_0}K_2$.

The upper bound of $C_{\texttt{key},i_p,j_p}^{(2)}$ is similar to \eqref{eq:upper_bound4} but with $M=2$ and  $\mathbf{H}_E$ known to Eve, i.e.,
\begin{align}\label{eq:upper_bound4b}
  &C_{s,i_p,j_p,\texttt{up}}^{(2)}\approx
  -h(\mathbf{Y}_E^{(2)}|\mathbf{H}_E)+h(\mathbf{Y}_{i_p}^{(2)},
  \mathbf{Y}_{E,p}^{(2)}|\mathbf{X}_{i_p},
  \mathbf{H}_{i_p,j_p},\mathbf{H}_E)\notag\\
  &\,\,+h(\mathbf{Y}_{j_p}^{(2)},\mathbf{Y}_{E,p}^{(2)}|\mathbf{X}_{j_p},
  \mathbf{H}_{j_p,i_p},\mathbf{H}_E)\notag\\
  &\,\,-h(\mathbf{Y}_{i_p}^{(2)},\mathbf{Y}_{j_p}^{(2)},\mathbf{Y}_{E,p}^{(2)}|
  \mathbf{X}_{i_p},\mathbf{X}_{j_p},
  \mathbf{H}_{i_p,j_p},\mathbf{H}_E)
\end{align}
where $\mathbf{H}_{i_p,j_p}=\mathbf{H}_{j_p,i_p}^T$. We know the DoF of each of the first three terms in \eqref{eq:upper_bound4b}. The last term in \eqref{eq:upper_bound4b} clearly has a zero DoF. One can verify that
  \begin{align}\label{eq:upper_bound4c}
  &\texttt{DoF}(C_{s,i_p,j_p,\texttt{up}}^{(2)})\notag\\
  &=-\min(N_E,N_{i_p}+N_{j_p})k_2+\min(N_E+N_{i_p},N_{j_p})k_2\notag\\
  &\,\,+\min(N_E+N_{j_p},N_{i_p})k_2.
\end{align}

The gap between the upper bound in \eqref{eq:upper_bound4c} and the lower bound in \eqref{eq:SDoF_pair_xx} is
\begin{equation}\label{}
  G_{i_p,j_p} = \left \{\begin{array}{cc}
                          0,& N_{i_p}\leq N_{j_p};\\
                          (\min(N_E+N_{j_p},N_{i_p})-N_{j_p})k_2,& N_{i_p}> N_{j_p}.
                        \end{array}
   \right .
\end{equation}
Note that the above implies that $\texttt{DoF}(C_{s,i_p,j_p,up}^{(2)})$ shown in \eqref{eq:upper_bound4c} is achievable by a public communication (for secret key generation) from the node with equal or smaller number of antennas to the other node.

Furthermore, if $N_i=N$ for all $i=1,\cdots,M$, then the pair-wise SDoF for phase 2 of the pair-wise ANECE with $M\geq 3$ is
 \begin{align}\label{eq:upper_bound4d}
  &\texttt{DoF}(C_{s,i_p,j_p,\texttt{up}}^{(2)})=\texttt{DoF}(C_{s,i_p,j_p,\texttt{low}}^{(2)})\notag\\
  &=(2N-\min(N_E,2N))k_2
\end{align}
where $k_2=\frac{K_2}{P_0}=\frac{2}{M(M-1)}K_2$. Here we see that the phase-2 pair-wise SDoF for the pair-wise ANECE decreases to zero rapidly as $\frac{1}{M^2}$ when the number $M$ of users in the network increases. Also for $N_E\geq 2N$, $\texttt{DoF}(C_{s,i_p,j_p,\texttt{up}}^{(2)})=0$.

\subsection{Summary}
\begin{Prop}\label{Property_Pairwise}
  The pair-wise ANECE is a scheme where two-user ANECE is applied to each pair of users in a network of $M\geq 3$ multi-antenna users within each coherence period, for which we have found:

1) For pair-wise ANECE, Eve of $N_E\geq 1$ antennas is able to determine her receive channel state information completely (at high power) relative to all users.

 2) While the phase-1 SDoF ($\texttt{SDoF}^{(1)}$) of pair-wise ANECE is the same as that for all-user ANECE, the pair-wise ANECE requires $\frac{M}{2}$ times more time-slots for phase-1 transmissions than all-user ANECE.

  3) The phase-2 SDoF ($\texttt{SDoF}^{(2)}$) of pair-wise ANECE is given by  \eqref{eq:upper_bound4c} regardless of $N_{i_p}\leq N_{j_p}$ or $N_{i_p}> N_{j_p}$.

   4) For a symmetric network where every node has the same number $N$ of antennas, $\texttt{SDoF}^{(2)}$ of pair-wise ANECE decreases to zero at least like $\frac{1}{M^2}$ as $M$ increases as shown in \eqref{eq:upper_bound4d}.

    5) For $M=3$ and $N_E\geq 2N$, $\texttt{SDoF}^{(2)}$ of pair-wise ANECE (see \eqref{eq:upper_bound4d}) is zero. This is in contrast to $\texttt{SDoF}^{(2)}$ of all-user ANECE (see \eqref{eq:69}) which is positive subject to $M=3$, $1\leq K_2<2N$ and (even) $N_E\geq 3N$.
\end{Prop}

\section{A Modified Two-User ANECE}\label{sec:general_ANECE}
We now focus on a network of two users with $N_1$ and $N_2$ antennas respectively against Eve with $N_E$ antennas. Assuming $N_1\leq N_2$, we consider what will be called a modified two-user ANECE where node 1 transmits $[\mathbf{P}_1,\mathbf{X}_1]\in\mathbb{C}^{N_1\times K}$ while node 2 transmits $[\mathbf{P}_2,\mathbf{X}_2]\in\mathbb{C}^{N_2\times K}$ over total $K$ slots in each coherence period, and each of $\mathbf{P}_1$ and $\mathbf{P}_2$ is a full-rank square matrix.

\begin{Assumption}\label{Assumption_P1P2}
  $\mathbf{P}_1$ has rank $N_1$ and dimension $N_1\times N_1$, and $\mathbf{P}_2$ has rank $N_2$ and  dimension $N_2\times N_2$.
\end{Assumption}

 Notice that the two square-shaped pilot matrices $\mathbf{P}_1$ and $\mathbf{P}_2$ have their lengths (numbers of columns) equal to $N_1$ and $N_2$ respectively. The two transmission phases of node 1 do not completely align with those of node 2 unless $N_1=N_2$.

We can ignore the propagation delays. Then
the signals received by node 1 during the slots $1\leq k\leq N_2$ can be described by
\begin{equation}\label{}
  \mathbf{Y}_1^{(1)}=\sigma\mathbf{H}_{1,2}\mathbf{P}_2+\mathbf{W}_1^{(1)},
\end{equation}
and the signals received by node 2 during  $1\leq k\leq N_1$ are
\begin{equation}\label{}
  \mathbf{Y}_2^{(1)}=\sigma\mathbf{H}_{2,1}\mathbf{P}_1+\mathbf{W}_2^{(1)}.
\end{equation}
Similarly, the signals received by node 1 during $N_2+1\leq k\leq K$ are
\begin{equation}\label{eq:Y1_2}
  \mathbf{Y}_1^{(2)}=\sigma\mathbf{H}_{1,2}\mathbf{X}_2+\mathbf{W}_1^{(2)},
\end{equation}
and the signals received by node 2 during $N_1+1\leq k\leq K$ are
\begin{equation}\label{eq:Y2_2}
  \mathbf{Y}_2^{(2)}=\sigma\mathbf{H}_{2,1}\mathbf{X}_1+\mathbf{W}_2^{(2)}.
\end{equation}

Subject to a high power in pilots, Node 1 can obtain its channel matrix $\mathbf{H}_{1,2}$ from $\mathbf{Y}_1^{(1)}$, and then receive the ``information'' in the $N_2\times (K-N_2)$ matrix $\mathbf{X}_2$ from $\mathbf{Y}_1^{(2)}$.
 Similarly, node 2 can obtain its channel matrix $\mathbf{H}_{2,1}$ from $\mathbf{Y}_2^{(1)}$, and then receive the ``information'' in the $N_1\times (K-N_1)$ matrix $\mathbf{X}_1$ from $\mathbf{Y}_2^{(2)}$. Note that we will still follow the generalized secrecy capacity. So the ``information'' above can be randomly generated by each node.

During $1\leq k\leq N_1$, Eve receives
\begin{equation}\label{eq:YE1'}
  \mathbf{Y}_E^{(1)'}=\sigma\mathbf{H}_E\mathbf{P}'
  +\mathbf{W}_E^{(1)'}
\end{equation}
with $\mathbf{H}_E=[\mathbf{H}_{E,1},\mathbf{H}_{E,2}]$, $\mathbf{P}'=[\mathbf{P}_1^H,\mathbf{P}_{2,1}^H]^H$ and $\mathbf{P}_{2,1}$ being the first $N_1$ columns of $\mathbf{P}_2$. Let $\mathbf{Q}_{P'}$ be a $(N_1+N_2)\times N_1$ orthonormal matrix spanning the range of $\mathbf{P}'$, and $\mathbf{Q}_{P',\perp}$ be a $(N_1+N_2)\times N_2$ orthogonal complement of $\mathbf{Q}_{P'}$. It follows from an analysis similar to \eqref{eq:YE1aa} that at high power Eve can uniquely determine $\mathbf{H}_{E,P'}\doteq \mathbf{H}_E\mathbf{Q}_{P'}$ from $\mathbf{Y}_E^{(1)'}$. But all entries of $\mathbf{H}_{E,P',\perp}\doteq \mathbf{H}_E\mathbf{Q}_{P',\perp}$ remain i.i.d. $\mathcal{CN}(0,1)$. Note that the dimensions of $\mathbf{H}_{E,P'}$ and $\mathbf{H}_{E,P',\perp}$ are $N_E\times N_1$ and $N_E\times N_2$ respectively.

During $N_1+1\leq k\leq K$, Eve receives
\begin{align}\label{eq:YE2P}
  &\mathbf{Y}_E^{(2)'}=\sigma\mathbf{H}_E\left [\begin{array}{cc}
                                       \mathbf{X}_{1,1} & \mathbf{X}_{1,2} \\
                                       \mathbf{P}_{2,2} & \mathbf{X}_2
                                     \end{array}
    \right ]
  +\mathbf{W}_E^{(2)'}\notag\\
  &=\sigma\mathbf{H}_E\mathbf{X}'
  +\mathbf{W}_E^{(2)'}
\end{align}
where $\mathbf{P}_{2,2}$ consists of the last $N_2-N_1$ columns of $\mathbf{P}_2$, and $[\mathbf{X}_{1,1}, \mathbf{X}_{1,2}]=\mathbf{X}_1$. Equivalently,
\begin{align}\label{eq:84}
  [\mathbf{Y}_E^{(1)'},\mathbf{Y}_E^{(2)'}]&=\sigma\mathbf{H}_E\left [ \begin{array}{ccc}
                                              \mathbf{P}_1 & \mathbf{X}_{1,1} & \mathbf{X}_{1,2} \\
                                              \mathbf{P}_{2,1} & \mathbf{P}_{2,2} & \mathbf{X}_2
                                            \end{array}
  \right ]\notag\\
  &\,\,+[\mathbf{W}_E^{(1)'},\mathbf{W}_E^{(2)'}].
\end{align}

Following a similar analysis shown in section \ref{sec:Phase_1}, one can verify that the DoF of $\mathbb{I}(\mathbf{Y}_1^{(1)};\mathbf{Y}_2^{(1)})$ is $N_1N_2$ subject to $\mathbf{H}_{1,2}=\mathbf{H}_{2,1}^T$, which is the SDoF between the users based on phase 2 of the modified two-user ANECE. This SDoF for phase 1 is no different from the pair-wise SDoF for phase 1 of the multi-user ANECE shown before.

The secrecy capacity for phase 2 of the modified two-user ANECE is now denoted by $C_{\texttt{key}}^{(2)}$, which satisfies
\begin{equation}\label{}
  C_{\texttt{key,low}}^{(2)}\leq C_{\texttt{key}}^{(2)}\leq C_{\texttt{key,up}}^{(2)}
\end{equation}
where
\begin{align}\label{eq:C_key_low_2}
  &C_{\texttt{key,low}}^{(2)}=\max_{i\neq j}(C_{\texttt{key},i,j,\texttt{low}}^{(2)})=C_{\texttt{key},0}^{(2)}-\min_{i\neq j} C_{i,E}'
\end{align}
with $i=1,2$ and $j=1,2$.
Furthermore,
\begin{equation}\label{}
  C_{\texttt{key},0}^{(2)}\approx \mathbb{I}(\mathbf{X}_1;\mathbf{Y}_2^{(2)}|\mathbf{H}_{2,1})+
  \mathbb{I}(\mathbf{X}_2;\mathbf{Y}_1^{(2)}|\mathbf{H}_{1,2}),
\end{equation}
  \begin{equation}\label{eq:CiEP}
    C_{i,E}'\approx \mathbb{I}(\mathbf{X}_i,\mathbf{Y}_i^{(2)};\mathbf{Y}_E^{(2)'}|\mathbf{H}_{i,j},
  \mathbf{H}_{E,P'}),
  \end{equation}
  \begin{equation}\label{eq:C_key_up}
    C_{\texttt{key,up}}^{(2)}\approx\mathbb{I}(\mathbf{X}_1,\mathbf{Y}_1^{(2)};\mathbf{X}_2,
    \mathbf{Y}_2^{(2)}|\mathcal{C}')
  \end{equation}
  where $\mathcal{C}'=\{\mathbf{H}_{1,2},\mathbf{H}_{2,1},\mathbf{H}_{E,P'},\mathbf{Y}_E^{(2)'}\}$.

  \subsection{A Remark on Channel-Model Based Secrecy Capacity}\label{sec:remark}
  From the perspective of wiretap channel model, another achievable secrecy capacity for phase 2 of the modified two-user ANECE (assuming a high power in phase 1) is
  \begin{align}\label{eq:C_low_2}
  &C_{\texttt{WTC,low}}^{(2)} =\mathbb{I}(\mathbf{X}_1;\mathbf{Y}_2^{(2)}|\mathbf{H}_{2,1})
  +\mathbb{I}(\mathbf{X}_2;\mathbf{Y}_1^{(2)}|\mathbf{H}_{1,2})-C_E'
  \end{align}
  where
  \begin{align}\label{}
   & C_E'=\mathbb{I}(\mathbf{X}_1,\mathbf{X}_2;\mathbf{Y}_E^{(1)'},\mathbf{Y}_E^{(2)'})
   =\mathbb{I}(\mathbf{X}_1,\mathbf{X}_2;\mathbf{Y}_E^{(2)'}|\mathbf{Y}_E^{(1)'})\notag\\
   &= \mathbb{I}(\mathbf{X}_1,\mathbf{X}_2;\mathbf{Y}_E^{(2)'}|\mathbf{H}_{E,P'}).
  \end{align}

  To compare $C_{\texttt{key,low}}^{(2)}$ with $C_{\texttt{WTC,low}}^{(2)}$, we only need to compare $C_{1,E}'$ and $C_{2,E}'$ defined in \eqref{eq:CiEP} with $C_E'$. For DoF analysis, we can replace the approximation at high power by equality. Let us first consider $C_{2,E}'$ for which we can write
  \begin{align}\label{}
    &C_{2,E}'= \mathbb{I}(\mathbf{X}_2,\mathbf{Y}_2^{(2)};\mathbf{Y}_E^{(2)'}|\mathbf{H}_{2,1},
  \mathbf{H}_{E,P'})\notag\\
  &=\mathbb{I}(\mathbf{X}_2,\mathbf{X}_1;\mathbf{Y}_E^{(2)'}|
  \mathbf{H}_{E,P'})=C_E'
  \end{align}
  where we have used the fact that $\mathbf{Y}_2^{(2)}=\sigma\mathbf{H}_{2,1}\mathbf{X}_1$ at high power and in this case $\mathbf{Y}_2^{(2)}$ and $\mathbf{X}_1$ imply each other when the $N_2\times N_1$ full column rank matrix $\mathbf{H}_{2,1}$ is given (here $N_1\leq N_2$).

  Next we consider $C_{1,E}'$ which is
   \begin{align}\label{}
    &C_{1,E}'= \mathbb{I}(\mathbf{X}_1,\mathbf{Y}_1^{(2)};\mathbf{Y}_E^{(2)'}|\mathbf{H}_{1,2},
  \mathbf{H}_{E,P'})\notag\\
  &\leq \mathbb{I}(\mathbf{X}_1,\mathbf{X}_2;\mathbf{Y}_E^{(2)'}|
  \mathbf{H}_{E,P'})=C_E'
  \end{align}
  where the inequality is due to the fact that at high power $\mathbf{Y}_1^{(2)}\approx\sigma\mathbf{H}_{1,2}\mathbf{X}_2$, and $\mathbf{Y}_1^{(2)}$ and $\mathbf{H}_{1,2}$ do not imply a unique $\mathbf{X}_2$ due to a right null space of the $N_1\times N_2$ matrix $\mathbf{H}_{1,2}$ for $N_1<N_2$.

  Therefore, it follows from \eqref{eq:C_key_low_2} and \eqref{eq:C_low_2} that for $N_1\leq N_2$,  $\texttt{DoF}(C_{\texttt{key,low}}^{(2)})\geq \texttt{DoF}(C_{\texttt{WTC,low}}^{(2)})$ where the strict inequality is expected if $N_1<N_2$.

  \subsection{Analysis of the Lower Bound $\max_{i\neq j}C_{\texttt{key},i,j,\texttt{low}}^{(2)}$}
  We will find next the DoF of $C_{\texttt{key},i,j,\texttt{low}}^{(2)}=C_{\texttt{key},0}^{(2)}-C'_{i,E}$. The DoF of the first term is simply
\begin{align}\label{eq:term1}
  &\texttt{DoF}(C_{\texttt{key},0}^{(2)}) = \texttt{DoF}(\mathbb{I}(\mathbf{X}_1;\mathbf{Y}_2^{(2)}|\mathbf{H}_{2,1}))\notag\\
  &+
  \texttt{DoF}(\mathbb{I}(\mathbf{X}_2;\mathbf{Y}_1^{(2)}|\mathbf{H}_{1,2}))\notag\\
  &= N_1(K-N_1)+N_1(K-N_2).
\end{align}
More effort is needed to find the DoF of $C'_{i,E}$.
Similar to \eqref{eq:CiE},
\begin{align}\label{eq:CiE_square}
  &C'_{i,E}\approx h(\mathbf{Y}_i^{(2)}|\mathbf{H}_{i,j})+h(\mathbf{Y}_E^{(2)'}|
  \mathbf{H}_{E,P'})\notag\\
  &\,\,-h(\mathbf{Y}_i^{(2)},\mathbf{Y}_E^{(2)'}|\mathbf{X}_i,\mathbf{H}_{i,j},
  \mathbf{H}_{E,P'}).
\end{align}
 It is clear that $\texttt{DoF}(h(\mathbf{Y}_1^{(2)}|\mathbf{H}_{1,2}))=N_1(K-N_2)$ and $\texttt{DoF}(h(\mathbf{Y}_2^{(2)}|\mathbf{H}_{2,1}))=N_1(K-N_1)$. The analyses of the other two terms in $C'_{i,E}$ are given below.

\subsection{Analysis of $h(\mathbf{Y}_E^{(2)'}|
  \mathbf{H}_{E,P'})$ in the Lower Bound}
  Refer to \eqref{eq:YE2P} and define the following $(\alpha,\beta)$-partitions
  \begin{equation}\label{}
   \mathbf{Y}_E^{(2)'}=[\mathbf{Y}_{E,\alpha},\mathbf{Y}_{E,\beta}]
  \end{equation}
  where $\mathbf{Y}_{E,\alpha}\in\mathbb{C}^{N_E\times \min(N_2,K-N_1)}$, $\mathbf{Y}_{E,\beta}\in\mathbb{C}^{N_E\times (K-N_T)^+}$. Namely,
  \begin{equation}\label{eq:YE2Pa}
  \mathbf{Y}_{E,\alpha}=\sigma\mathbf{H}_E\mathbf{X}_\alpha
  +\mathbf{W}_{E,\alpha},
\end{equation}
 \begin{equation}\label{eq:YE2Pb}
  \mathbf{Y}_{E,\beta}=\sigma\mathbf{H}_E\mathbf{X}_\beta
  +\mathbf{W}_{E,\beta}.
\end{equation}
where $[\mathbf{X}_\alpha,\mathbf{X}_\beta]=\mathbf{X}'$, $\mathbf{X}_\alpha\in\mathbb{C}^{N_T\times \min(N_2,K-N_1)}$ and $\mathbf{X}_\beta\in\mathbb{C}^{N_T\times (K-N_T)^+}$.

It follows that due to the unknown $\mathbf{H}_{E,P',\perp}\in\mathbb{C}^{N_E\times N_2}$,
\begin{align}\label{eq:DoFhYEalpha}
&\texttt{DoF}(h(\mathbf{Y}_{E,\alpha}|
  \mathbf{H}_{E,P'}))\geq \texttt{DoF}(h(\mathbf{Y}_{E,\alpha}|\mathbf{X}_\alpha,
  \mathbf{H}_{E,P'}))\notag\\
  &=N_E\min(N_2,K-N_1).
\end{align}
Since this is the total number of entries in $\mathbf{Y}_{E,\alpha}$, the equality in the above ``$\geq$'' holds.

If $K>N_T$, we define
\begin{align}\label{eq:YEP}
  &\mathbf{Y}_{E,P}\doteq[\mathbf{Y}_E^{(1)'},\mathbf{Y}_{E,\alpha}]
  =\sigma\mathbf{H}_E\mathbf{P}_T+\mathbf{W}_{E,P}
\end{align}
with $\mathbf{P}_T=[\mathbf{P}',\mathbf{X}_\alpha]\in\mathbb{C}^{N_T\times N_T}$. At high power, $\mathbf{H}_E\approx \frac{1}{\sigma}\mathbf{Y}_{E,P}\mathbf{P}_T^{-1}$, which however depends on $\mathbf{X}_\alpha$.

Let the $(a,b)$-partitions of $\mathbf{Y}_{E,\beta}$ be
\begin{equation}\label{}
  \mathbf{Y}_{E,\beta}=[\mathbf{Y}_{E,\beta,a}^T,\mathbf{Y}_{E,\beta,b}^T]^T
\end{equation}
 with $\mathbf{Y}_{E,\beta,a}\in\mathbb{C}^{\min(N_E,N_T)\times (K-N_T)^+}$ and $\mathbf{Y}_{E,\beta,b}\in\mathbb{C}^{(N_E-N_T)^+\times (K-N_T)^+}$.

It follows that due to the unknown $\mathbf{X}_\beta$ (consisting of i.i.d. entries),
\begin{align}\label{eq:DoFhYEBa}
&\texttt{DoF}(h(\mathbf{Y}_{E,\beta,a}|\mathbf{Y}_{E,\alpha},
  \mathbf{H}_{E,P'}))\notag\\
  &\geq \texttt{DoF}(h(\mathbf{Y}_{E,\beta,a}|\mathbf{Y}_{E,\alpha},
  \mathbf{H}_{E,P'},\mathbf{X}_\alpha))\notag\\
  &=\texttt{DoF}(h(\mathbf{Y}_{E,\beta,a}|\mathbf{Y}_{E,P},
  \mathbf{P}_T))\notag\\
  &=\texttt{DoF}(h(\mathbf{Y}_{E,\beta,a}|\mathbf{H}_E))\notag\\
  &
  =\min(N_E,N_T)(K-N_T)^+.
\end{align}
Note that conditioning on $\mathbf{H}_{E,P'}$ is equivalent at high power to conditioning on $\mathbf{Y}_E^{(1)'}$ in terms of DoF.
Also, the equality in the above ``$\geq$'' holds since $\mathbf{Y}_{E,\beta,a}$ has total $\min(N_E,N_T)(K-N_T)^+$ elements.

If $K>N_T$ and $N_E>N_T$, then $\mathbf{Y}_{E,\beta,b}$ exists. In this case, we have at high power that $\mathbf{H}_{E,a}\approx \frac{1}{\sigma}\mathbf{Y}_{E,P,a}\mathbf{P}_T^{-1}$ and $\mathbf{H}_{E,b}\approx \frac{1}{\sigma}\mathbf{Y}_{E,P,b}\mathbf{P}_T^{-1}$. Also $\mathbf{Y}_{E,\beta,a}\approx \sigma \mathbf{H}_{E,a}\mathbf{X}_\beta$ implies $\mathbf{X}_\beta
=\frac{1}{\sigma}\mathbf{H}_{E,a}^{-1}\mathbf{Y}_{E,\beta,a}$ where $\mathbf{H}_{E,a}^{-1}\in\mathbb{C}^{N_T\times N_T}$ exists with probability one. Finally,
\begin{align}\label{}
  &\mathbf{Y}_{E,\beta,b}\approx \sigma \mathbf{H}_{E,b}\mathbf{X}_\beta
  \approx \mathbf{H}_{E,b}\mathbf{H}_{E,a}^{-1}\mathbf{Y}_{E,\beta,a}\notag\\
  &
  \approx \mathbf{Y}_{E,P,b}\mathbf{Y}_{E,P,a}^{-1}\mathbf{Y}_{E,\beta,a}.
\end{align}
Therefore,
\begin{equation}\label{eq:DoFhYEBb}
  \texttt{DoF}(h(\mathbf{Y}_{E,\beta,b}|\mathbf{Y}_{E,\beta,a},\mathbf{Y}_{E,\alpha},
  \mathbf{H}_{E,P'}))=0.
\end{equation}

Adding \eqref{eq:DoFhYEalpha}, \eqref{eq:DoFhYEBa} and \eqref{eq:DoFhYEBb} yields
\begin{align}\label{eq:term2}
&\texttt{DoF}(h(\mathbf{Y}_E^{(2)'}|
  \mathbf{H}_{E,P'}))=
  N_E\min(N_2,K-N_1)\notag\\
  &\,\,+\min(N_E,N_T)(K-N_T)^+.
\end{align}

\subsection{Analysis of $h(\mathbf{Y}_i^{(2)},\mathbf{Y}_E^{(2)'}|\mathbf{X}_i,\mathbf{H}_{i,j},
  \mathbf{H}_{E,P'})$ in the Lower Bound}
  \subsubsection{For $h(\mathbf{Y}_1^{(2)},\mathbf{Y}_E^{(2)'}|\mathbf{X}_1,\mathbf{H}_{1,2},
  \mathbf{H}_{E,P'})$}
  We will write it as $h(\mathbf{Y}_1^{(2)},\mathbf{Y}_E^{(2)'}|\mathcal{C}_{1,2})$ with $\mathcal{C}_{1,2}=\{\mathbf{X}_1,\mathbf{H}_{1,2},
  \mathbf{H}_{E,P'}\}$. Here $\mathbf{X}_1$ is among the conditions. It follows from \eqref{eq:Y1_2} that $h(\mathbf{Y}_1^{(2)}|\mathcal{C}_{1,2})=
  h(\mathbf{Y}_1^{(2)}|\mathbf{H}_{1,2})$ and hence
  \begin{align}\label{eq:low1}
    &\texttt{DoF}(h(\mathbf{Y}_1^{(2)}|\mathcal{C}_{1,2}))=N_1(K-N_2).
  \end{align}
  Next we need to consider the second component $h(\mathbf{Y}_E^{(2)'}|\mathbf{Y}_1^{(2)},\mathcal{C}_{1,2})$. Conditioned on $\mathbf{Y}_1^{(2)}$ and $\mathcal{C}_{1,2}$ at high power, we know from \eqref{eq:Y1_2} that $\mathbf{X}_2\approx\mathbf{X}_{2,0}+\mathbf{N}_{x2}\mathbf{T}_{x2}$ where $\mathbf{X}_{2,0}$ is the minimum norm solution of $\mathbf{X}_2$ to $\mathbf{Y}_1^{(2)}=\sigma\mathbf{H}_{1,2}\mathbf{X}_2$, the columns of $\mathbf{N}_{x2}\in\mathbb{C}^{N_2\times \Delta N}$ span the right null space of $\mathbf{H}_{1,2}$, and $\mathbf{T}_{x2}\in \mathbb{C}^{\Delta N\times (K-N_2)}$ is arbitrary. Then, \eqref{eq:YE2P} becomes
  \begin{equation}\label{eq:YE2P2}
  \mathbf{Y}_E^{(2)'}=\sigma\mathbf{H}_E\left [\begin{array}{cc}
                                       \mathbf{X}_{1,1} & \mathbf{X}_{1,2} \\
                                       \mathbf{P}_{2,2} & \mathbf{X}_{2,0}+\mathbf{N}_{x2}\mathbf{T}_{x2}
                                     \end{array}
    \right ]
  +\mathbf{W}_E^{(2)'}
\end{equation}
where only $\mathbf{T}_{x2}$ and $\mathbf{H}_{E,P',\perp}$ (inside $\mathbf{H}_E=[\mathbf{H}_{E,P'},\mathbf{H}_{E,P',\perp}]\mathbf{Q}^H$) are unknown subject to given $\{\mathbf{Y}_1^{(2)},\mathcal{C}_{1,2}\}$ and high power.

Recall the $(\alpha,\beta)$-partitions $\mathbf{Y}_E^{(2)'}=[\mathbf{Y}_{E,\alpha},\mathbf{Y}_{E,\beta}]$
with $\mathbf{Y}_{E,\alpha}\in\mathbb{C}^{N_E\times \min(N_2,K-N_1)}$ and $\mathbf{Y}_{E,\beta}\in\mathbb{C}^{N_E\times (K-N_T)^+}$.
Also let $\mathbf{T}_{x2}=[\mathbf{T}_{x2,\alpha'},\mathbf{T}_{x2,\beta}]$ with $\mathbf{T}_{x2,\alpha'}\in\mathbb{C}^{\Delta N\times \min(N_1,K-N_2)}$ and $\mathbf{T}_{x2,\beta}\in\mathbb{C}^{\Delta N\times (K-N_T)^+}$. Note that $\mathbf{Y}_{E,\alpha}$ depends on $\mathbf{T}_{x2,\alpha'}$, and $\mathbf{Y}_{E,\beta}$ depends on $\mathbf{T}_{x2,\beta}$.

Following a similar analysis leading to \eqref{eq:DoFhYEalpha}, we have that due to unknown $\mathbf{H}_{E,P',\perp}\in\mathbb{C}^{N_E\times N_2}$,
\begin{align}\label{eq:low2}
  &\texttt{DoF}(h(\mathbf{Y}_{E,\alpha}|\mathbf{Y}_1^{(2)},\mathcal{C}_{1,2}))
  =\texttt{DoF}(h(\mathbf{Y}_{E,\alpha}|\mathcal{C}_{1,2}))\notag\\
  &=N_E\min(N_2,K-N_1)
\end{align}
which holds with or without the knowledge of $\mathbf{T}_{x2,\alpha'}$ due to the ``fullness'' of DoF (i.e., the above DoF equals to the number of entries in $\mathbf{Y}_{E,\alpha}$ even when conditioned on $\mathbf{T}_{x2,\alpha'}$).

For $K>N_T$, given $\mathbf{Y}_{E,\alpha}$ along with $\{\mathbf{Y}_1^{(2)},\mathcal{C}_{1,2}\}$, we know $\mathbf{H}_E$ at high power if $\mathbf{T}_{x2,\alpha'}$ is also given.

Also for $K>N_T$, we now use the following $(a',b')$-partitions
\begin{equation}\label{}
  \mathbf{Y}_{E,\beta}=[\mathbf{Y}_{E,\beta,a'}^T,\mathbf{Y}_{E,\beta,b'}^T]^T
\end{equation}
 with $\mathbf{Y}_{E,\beta,a'}\in\mathbb{C}^{\min(N_E,\Delta N)\times (K-N_T)^+}$ and $\mathbf{Y}_{E,\beta,b'}\in\mathbb{C}^{(N_E-\Delta N)^+\times (K-N_T)^+}$. It follows that
due to unknown $\mathbf{T}_{x2,\beta}\in \mathbb{C}^{\Delta N\times (K-N_T)^+}$,
\begin{align}\label{eq:low3}
  &\texttt{DoF}(h(\mathbf{Y}_{E,\beta,a'}|\mathbf{Y}_{E,\alpha},
  \mathbf{Y}_1^{(2)},\mathcal{C}_{1,2}))\notag\\
  &\geq \texttt{DoF}(h(\mathbf{Y}_{E,\beta,a'}|\mathbf{T}_{x2,\alpha'},\mathbf{Y}_{E,\alpha},
  \mathbf{Y}_1^{(2)},\mathcal{C}_{1,2}))\notag\\
  &=\texttt{DoF}(h(\mathbf{Y}_{E,\beta,a'}|\mathbf{H}_E,\mathbf{X}_1,\mathbf{X}_{2,0},\mathbf{N}_{x2}))
  \notag\\
  &=\min(N_E,\Delta N)(K-N_T)^+
\end{align}
where the equality in ``$\geq$'' actually holds due to the fullness of DoF.

Next we show that for $K>N_T$ and $N_E>\Delta N$, $\mathbf{Y}_{E,\beta,b'}$ is determined with probability one at high power by $\{\mathbf{Y}_{E,\beta,a'}, \mathbf{Y}_{E,\alpha},
  \mathbf{Y}_1^{(2)},\mathcal{C}_{1,2}\}$.

With given $\mathbf{H}_{E,P'}$ (or equivalently $\mathbf{Y}_E^{(1)'}$) and the first $\Delta N$ columns of $\mathbf{Y}_{E,\alpha}$, we know that $\mathbf{H}_E\approx \mathbf{\bar H}_E+\mathbf{T}_E\mathbf{N}_E$ where only $\mathbf{T}_E\in\mathbb{C}^{N_E\times N_1}$ is unknown.  Note that the row span of $\mathbf{N}_E$ is the left null space of $\mathbf{P}''\doteq \left [\begin{array}{cc}
                                      \mathbf{P}_1& \mathbf{X}_{1,1}  \\
                                      \mathbf{P}_{2,1} &\mathbf{P}_{2,2}
                                     \end{array}
    \right ]$. Let $\mathbf{Y}_{E,\alpha'}$ be the last $N_1$ columns of $\mathbf{Y}_{E,\alpha}$. It follows that
\begin{equation}\label{}
  \mathbf{Y}_{E,\alpha'}\approx \sigma(\mathbf{\bar H}_E+\mathbf{T}_E\mathbf{N}_E)
  \left [\begin{array}{c}
                                       \mathbf{X}_{1,2,\alpha'} \\
                                        \mathbf{X}_{2,0,\alpha'}+
                                        \mathbf{N}_{x2}\mathbf{T}_{x2,\alpha'}
                                     \end{array}
    \right ]
\end{equation}
where $\mathbf{T}_E\in\mathbb{C}^{N_E\times N_1}$ and $\mathbf{T}_{x2,\alpha'}\in\mathbb{C}^{\Delta N\times N_1}$ are the only (inter-dependent) unknowns. Specifically, $\mathbf{T}_E$ is a (uniquely valued) function of the unknown $\mathbf{T}_{x2,\alpha'}$ subject to given $\{\mathbf{Y}_{E,\alpha},
  \mathbf{Y}_1^{(2)},\mathcal{C}_{1,2}\}$. This is because of a property similar to that of $\mathbf{H}_E$ in \eqref{eq:YEP}.

Furthermore, we can write the $(a',b')$-partitions of $\mathbf{Y}_{E,\beta}$ as follows:
\begin{equation}\label{}
  \mathbf{Y}_{E,\beta,a'}\approx \sigma(\mathbf{\bar H}_{E,a'}+\mathbf{T}_{E,a'}\mathbf{N}_E)
  \mathbf{X}_{2,\beta}',
\end{equation}
\begin{equation}\label{}
  \mathbf{Y}_{E,\beta,b'}\approx \sigma(\mathbf{\bar H}_{E,b'}+\mathbf{T}_{E,b'}\mathbf{N}_E)
  \mathbf{X}_{2,\beta}',
\end{equation}
where
\begin{equation}\label{}
  \mathbf{X}_{2,\beta}'=\left [\begin{array}{c}
                                       \mathbf{X}_{1,2,\beta} \\
                                        \mathbf{X}_{2,0,\beta}+
                                        \mathbf{N}_{x2}\mathbf{T}_{x2,\beta}
                                     \end{array}
    \right ].
\end{equation}
We see that subject to $\mathbf{Y}_{E,\beta,a'}$ in addition to  $\{\mathbf{Y}_{E,\alpha},\mathbf{Y}_1^{(2)},\mathcal{C}_{1,2}\}$, $\mathbf{T}_{E,a'}$ (and hence $\mathbf{T}_E$) is also a function of the unknown $\mathbf{T}_{x2,\beta}$. This makes $\mathbf{T}_E$ deterministic with probability one after $\{\mathbf{Y}_{E,\beta,a'},\mathbf{Y}_{E,\alpha},\mathbf{Y}_1^{(2)},\mathcal{C}_{1,2}\}$ is given at high power.
The above condition hence determines a unique $\mathbf{H}_E$, hence a unique  $\mathbf{T}_{x2,\beta}$ and hence a unique $\mathbf{Y}_{E,\beta,b'}$. Namely,
\begin{align}\label{eq:low4}
  &\texttt{DoF}(h(\mathbf{Y}_{E,\beta,b'}|\mathbf{Y}_{E,\beta,a'},\mathbf{Y}_{E,\alpha},
  \mathbf{Y}_1^{(2)},\mathcal{C}_{1,2}))=0.
\end{align}

  Therefore, adding \eqref{eq:low1}, \eqref{eq:low2}, \eqref{eq:low3} and \eqref{eq:low4} yields
  \begin{align}\label{eq:term3}
  &\texttt{DoF}(h(\mathbf{Y}_1^{(2)},\mathbf{Y}_E^{(2)'}|\mathcal{C}_{1,2}))= N_1(K-N_2)\notag\\
  &\,\,+N_E\min(N_2,K-N_1)+\min(N_E,\Delta N)(K-N_T)^+.
  \end{align}

  \subsubsection{For $h(\mathbf{Y}_2^{(2)},\mathbf{Y}_E^{(2)'}|\mathbf{X}_2,\mathbf{H}_{2,1},
  \mathbf{H}_{E,P'})$} Let $\mathcal{C}_{2,1}=\{\mathbf{X}_2,\mathbf{H}_{2,1},
  \mathbf{H}_{E,P'}\}$. It follows from \eqref{eq:Y1_2} that
  \begin{align}\label{}
    &\texttt{DoF}(h(\mathbf{Y}_2^{(2)}|\mathcal{C}_{2,1}))
    =\texttt{DoF}(h(\mathbf{Y}_2^{(2)}|\mathbf{H}_{2,1}))\notag\\
    &=N_1(K-N_1).
  \end{align}
  Given $\mathbf{Y}_2^{(2)}$ and $\mathbf{H}_{2,1}$, $\mathbf{X}_1$ is unique (because of $N_2\geq N_1$) at high power, and in this case, only $\mathbf{H}_{E,P',\perp}$  inside $\mathbf{H}_E$ shown in \eqref{eq:YE2P} is unknown subject to given $\mathcal{C}_{2,1}$ and high power. Recalling $\mathbf{Y}_E^{(2)'}=[\mathbf{Y}_{E,\alpha},\mathbf{Y}_{E,\beta}]$, it follows that
  \begin{align}\label{}
    &\texttt{DoF}(h(\mathbf{Y}_{E,\alpha}|\mathbf{Y}_2^{(2)},\mathcal{C}_{2,1}))
    =\texttt{DoF}(h(\mathbf{Y}_{E,\alpha}|\mathbf{X}_1,\mathbf{X}_2,
  \mathbf{H}_{E,P'}))\notag\\
  &=N_E\min(N_2,K-N_1),
  \end{align}
  and for $K>N_T$,
 \begin{align}
 &\texttt{DoF}(h(\mathbf{Y}_{E,\beta}|\mathbf{Y}_{E,\alpha},
    \mathbf{Y}_2^{(2)},\mathcal{C}_{2,1}))\notag\\
 & =\texttt{DoF}(h(\mathbf{Y}_{E,\beta}|\mathbf{X}_1,\mathbf{X}_2,
  \mathbf{H}_E))=0.
  \end{align}

  Therefore,
  \begin{align}\label{eq:term4}
  &\texttt{DoF}(h(\mathbf{Y}_2^{(2)},\mathbf{Y}_E^{(2)'}|\mathcal{C}_{2,1}))=N_1(K-N_1)
  \notag\\
  &\,\,+N_E\min(N_2,K-N_1).
  \end{align}

\subsection{Summary of the Lower Bound}
Combining the expressions in \eqref{eq:term1}, \eqref{eq:CiE_square}, \eqref{eq:term2} and \eqref{eq:term3}, we have
\begin{align}\label{eq:DoF_C_key}
&\texttt{SDoF}_{\texttt{new}}^{(2)}\doteq \texttt{DoF}(C_{\texttt{key},1,2,\texttt{low}}^{(2)})\notag\\
&= N_1(2K-N_T)+\min(N_E,\Delta N)(K-N_T)^+\notag\\
&\,\,-\min(N_E,N_T)(K-N_T)^+\notag\\
&=\left \{\begin{array}{cc}
            N_1(2K-N_T), & \texttt{C1}; \\
            N_1(2K-N_T)-(N_E-\Delta N)(K-N_T)^+, & \texttt{C2}; \\
            N_1[2K-N_T-(2K-2N_T)^+], & \texttt{C3};
          \end{array}
 \right .
\end{align}
where
 $K\geq N_2\geq N_1$, $N_T=N_1+N_2$, $\Delta N=N_2-N_1$, and ``\texttt{C1}, \texttt{C2}, \texttt{C3}'' specify three (small, medium and large) regions of $N_E$ as defined below \eqref{eq:SDoF_upB2}.

We see that $\texttt{SDoF}_{\texttt{new}}^{(2)}$ is non-increasing function of $N_E$ and non-decreasing function of $K$, which is consistent with intuitions. In the small and medium regions \texttt{C1} and \texttt{C2} of $N_E$ (i.e., $0\leq N_E<N_T$ specifically), $\texttt{SDoF}_{\texttt{new}}^{(2)}$ increases with $K$. But for the large region \texttt{C3} of $N_E$ (i.e., $N_E\geq N_T$), $\texttt{SDoF}_{\texttt{new}}^{(2)}$ increases with $K$ for $N_2\leq K\leq N_T$ and stays constant of $K$ for $K\geq N_T$. For large $N_E$ and $K$ (i.e., $N_E\geq N_T$ and $K\geq N_T$), $\texttt{SDoF}_{\texttt{new}}^{(2)}=N_1N_T$.

Combining the expressions in \eqref{eq:term1}, \eqref{eq:CiE_square}, \eqref{eq:term2} and \eqref{eq:term4}, we have
\begin{align}
&\texttt{DoF}(C_{\texttt{key},2,1,\texttt{low}}^{(2)})=N_1(2K-N_T)\notag\\
&\,\,-\min(N_E,N_T)(K-N_T)^+.
\end{align}

It is clear that
\begin{align}
&\texttt{DoF}(C_{\texttt{key},1,2,\texttt{low}}^{(2)})-\texttt{DoF}(C_{\texttt{key},2,1,\texttt{low}}^{(2)})\notag\\
&\geq \min(N_E,\Delta N)(K-N_T)^+
\geq 0.
\end{align}

\subsection{Analysis of the Upper Bound}
We now analyze $C_{\texttt{key,up}}^{(2)}$ in \eqref{eq:C_key_up}.
Similar to \eqref{eq:upper_bound2} and \eqref{eq:upper_bound4}, we can write
\begin{align}\label{eq:upper_bound4new}
  &C_{\texttt{key,up}}^{(2)}\approx
  -h(\mathbf{Y}_E^{(2)'}|\mathbf{H}_{E,P'})\notag\\
  &\,\,+h(\mathbf{Y}_1^{(2)},\mathbf{Y}_E^{(2)'}|\mathbf{X}_1,
  \mathbf{H}_{1,2},\mathbf{H}_{E,P'})\notag\\
  &\,\,+h(\mathbf{Y}_2^{(2)},\mathbf{Y}_E^{(2)'}|\mathbf{X}_2,
  \mathbf{H}_{2,1},\mathbf{H}_{E,P'})\notag\\
  &\,\,-h(\mathbf{Y}_1^{(2)},\mathbf{Y}_2^{(2)},\mathbf{Y}_E^{(2)'}|\mathbf{X}_1,\mathbf{X}_2,
  \mathbf{H}_{1,2},\mathbf{H}_{2,1},\mathbf{H}_{E,P'}).
\end{align}
The DoF of the first three terms are given by \eqref{eq:term2}, \eqref{eq:term3} and \eqref{eq:term4}. For the last term, we let $\mathcal{C}_T=\{\mathbf{X}_1,\mathbf{X}_2,
  \mathbf{H}_{1,2},\mathbf{H}_{2,1},\mathbf{H}_{E,P'})\}$. Then, we see $\texttt{DoF}(h(\mathbf{Y}_1^{(2)}|\mathcal{C}_T))=0$, $\texttt{DoF}(h(\mathbf{Y}_2^{(2)}|\mathbf{Y}_1^{(2)},\mathcal{C}_T))=0$, and hence $\texttt{DoF}(h(\mathbf{Y}_1^{(2)},\mathbf{Y}_2^{(2)},\mathbf{Y}_E^{(2)'}|\mathcal{C}_T))=
  \texttt{DoF}(h(\mathbf{Y}_E^{(2)'}|\mathcal{C}_T))=
  \texttt{DoF}(h(\mathbf{Y}_{E,\alpha}|\mathcal{C}_T))+\texttt{DoF}(h(\mathbf{Y}_{E,\beta}|
  \mathbf{Y}_{E,\alpha},
  \mathcal{C}_T))=
  N_E\Delta N$
  where we have used $\texttt{DoF}(h(\mathbf{Y}_{E,\beta}|
  \mathbf{Y}_{E,\alpha},
  \mathcal{C}_T))=\texttt{DoF}(h(\mathbf{Y}_{E,\beta}|
  \mathbf{H}_E,\mathbf{X}_1,\mathbf{X}_2))=0$.

  Combining the above results, one can verify that
  \begin{equation}\label{eq:up_and_low}
    \texttt{DoF}(C_{\texttt{key,up}}^{(2)})=\texttt{DoF}(C_{\texttt{key},1,2,\texttt{low}}^{(2)})
  \end{equation}
  which is given in \eqref{eq:DoF_C_key}. Therefore, $\texttt{DoF}(C_{\texttt{key}}^{(2)})$ is also given by \eqref{eq:DoF_C_key}.

  \subsection{Summary}
\begin{Prop}\label{Property_two_users} For a network of two users with $N_1\leq N_2$, $N_T=N_1+N_2$, and $\Delta N=N_2-N_1$, we have found:

1)  The lower and upper bounds on the phase-2 SDoF of the modified two-user ANECE coincide.

2) The total SDoF of the modified two-user ANECE equals $N_1N_2+\texttt{DoF}(C_{\texttt{key}}^{(2)})$ where $\texttt{DoF}(C_{\texttt{key}}^{(2)})$ is given by \eqref{eq:up_and_low} or equivalently \eqref{eq:DoF_C_key}.

3)  For $N_E<N_T$, $\texttt{SDoF}_{\texttt{new}}^{(2)}$ increases linearly with $K$.

4) For $N_E\geq N_T$, $\texttt{SDoF}_{\texttt{new}}^{(2)}$ increases linearly with $K$ for $N_2\leq K\leq N_T$, and equals $N_1N_T$ for $K\geq N_T$.
\end{Prop}

\section{Comparison of the Original Two-User ANECE and the Modified Two-User ANECE}\label{sec:Comparison}
The phase-1 SDoF of the two versions is the same, i.e., $N_1N_2$. We will only need to consider the phase-2 SDoF while assuming $N_1<N_2$. (The two versions would be identical if $N_1=N_2$.)

The phase-2 SDoF of the original version is given by \eqref{eq:SDoF_upB2} while that of the modified version is given by \eqref{eq:DoF_C_key}. For a fair comparison, we will let $K_2=K-N_2>0$. It is easy to verify that
\begin{align}\label{}
  &\texttt{SDoF}_{\texttt{new}}^{(2)}-\texttt{SDoF}_{\texttt{original}}^{(2)}
  =N_1(N_2-N_1)
\end{align}
for all cases of $N_E\geq 0$ and $K\geq N_2$. It is interesting to observe that this gap equals the number of entries in $\mathbf{X}_{1,1}$ in the modified two-user ANECE. See $\mathbf{X}_{1,1}$ in \eqref{eq:84} for example.

\begin{Prop}\label{Property_two_users_comparison}
  The phase-2 SDoF of the modified two-user ANECE (see \eqref{eq:DoF_C_key}) is $N_1\Delta N$ larger than that (see \eqref{eq:SDoF_upB2}) of the original two-user ANECE, subject to $K_2=K-N_2\geq 0$.
\end{Prop}

\section{Conclusion}\label{sec:conclusion}
We have shown novel insights into the (pair-wise) SDoF of a multi-user multi-antenna full-duplex wireless network using multi-user multi-antenna ANECE against Eve with any number of antennas. These insights are based on the DoF of lower and upper bounds on the pair-wise SKC between users using the observations from the multi-user ANECE. In many cases, the lower and upper bounds have the same DoF, which hence reveal the exact SDoF. In particular, the obtained lower and upper bounds on SDoF for two-user ANECE coincide, regardless of the numbers of antennas on users and Eve, and regardless of the number of time slots used for phase 2 in each coherence period. These SDoF bounds also coincide for multi-user ANECE when every user has the same number of antennas and the number of time slots used for phase 2 is no larger than the number of antennas. And for a three-user network, all-user ANECE has been shown to have advantages over pair-wise ANECE in both phases 1 and 2.

We have also shown that when two users have different numbers of antennas, a modified two-user ANECE, for which each user applies a square-shaped nonsingular pilot matrix, has a larger SDoF than the original two-user ANECE for which the pilot matrices from both users consume the same number of time-slots. The proven result is also explainable intuitively.

The lower and upper bounds on the pair-wise SKC used in this paper are the same as those established by Maurer \cite{Maurer1993} and others \cite{Bloch2011}. The matching DoF of the lower and upper bounds in many cases as shown in this paper is a pleasant discovery. A similar phenomenon was also found recently in \cite{Hua2023} for two-user half-duplex schemes.

This paper has given a more complete picture of the pair-wise SDoF of multi-user ANECE. But many questions are still open, which include how much correlation there is between the keys generated by different pairs of users for phase 2, how undesirable this potential correlation could be in practice, and whether there is another multi-user scheme that can yield a higher SDoF. Even for the two-user case, it is not yet formally proven whether the modified two-user ANECE is already optimal among all possible schemes in terms of SDoF.
We hope that good answers to the above questions will be found in the near future.

\appendix

\subsection{Analysis of $C_{\texttt{key},i,j,\texttt{up}}^{(2)}$ in \eqref{eq:upper_bound}}\label{sec:upper_bound}
In this section, we analyse the DoF of $C_{\texttt{key},i,j,\texttt{up}}^{(2)}$ shown in \eqref{eq:upper_bound}. It follows from \eqref{eq:upper_bound}
 that with a high power in phase 1,
\begin{align}\label{eq:upper_bound2}
  &C_{\texttt{key},i,j,\texttt{up}}^{(2)}\approx \mathbb{I}(\mathbf{X}_i,\mathbf{Y}_i^{(2)};\mathbf{X}_j,\mathbf{Y}_j^{(2)}|
  \mathcal{C})
\end{align}
with $\mathcal{C}=\{\mathbf{H}_i,\mathbf{H}_j,\mathbf{H}_{E,P},\mathbf{Y}_E^{(2)}\}$. Furthermore,
\begin{align}\label{eq:upper_bound3}
  &C_{\texttt{key},i,j,\texttt{up}}^{(2)}\approx h(\mathbf{X}_i,\mathbf{Y}_i^{(2)}|
  \mathcal{C})-h(\mathbf{X}_i,\mathbf{Y}_i^{(2)}|\mathbf{X}_j,\mathbf{Y}_j^{(2)},
  \mathcal{C})\notag\\
  &=h(\mathbf{X}_i|
  \mathcal{C})+h(\mathbf{Y}_i^{(2)}|\mathbf{X}_i,
  \mathcal{C})\notag\\
  &\,\,-h(\mathbf{X}_i|\mathbf{X}_j,\mathbf{Y}_j^{(2)},
  \mathcal{C})-h(\mathbf{Y}_i^{(2)}|\mathbf{X}_i,\mathbf{X}_j,\mathbf{Y}_j^{(2)},
  \mathcal{C}).
\end{align}
Using $h(A|B,C)=h(A,B|C)-h(B|C)=h(B|A,C)+h(A|C)-h(B|C)$, each of the four terms in the above can be rewritten as follows:
\begin{align}\label{eq:t1}
&h(\mathbf{X}_i|
  \mathcal{C})=h(\mathbf{X}_i|
  \mathbf{H}_{E,P},\mathbf{Y}_E^{(2)})=h(\mathbf{Y}_E^{(2)}|\mathbf{X}_i,\mathbf{H}_{E,P})\notag\\
  &
  \,\,+h(\mathbf{X}_i|\mathbf{H}_{E,P})-h(\mathbf{Y}_E^{(2)}|\mathbf{H}_{E,P})
  \notag\\
  &=h(\mathbf{Y}_E^{(2)}|\mathbf{X}_i,\mathbf{H}_{E,P})+h(\mathbf{X}_i)
  -h(\mathbf{Y}_E^{(2)}|\mathbf{H}_{E,P}),
\end{align}
\begin{align}\label{eq:t2}
&h(\mathbf{Y}_i^{(2)}|\mathbf{X}_i,
  \mathcal{C})=h(\mathbf{Y}_i^{(2)}|\mathbf{X}_i,
  \mathbf{H}_i,\mathbf{H}_{E,P},\mathbf{Y}_E^{(2)})\notag\\
  &=h(\mathbf{Y}_i^{(2)},\mathbf{Y}_E^{(2)}|\mathbf{X}_i,
  \mathbf{H}_i,\mathbf{H}_{E,P})-h(\mathbf{Y}_E^{(2)}|\mathbf{X}_i,
  \mathbf{H}_i,\mathbf{H}_{E,P})\notag\\
  &=h(\mathbf{Y}_i^{(2)},\mathbf{Y}_E^{(2)}|\mathbf{X}_i,
  \mathbf{H}_i,\mathbf{H}_{E,P})-h(\mathbf{Y}_E^{(2)}|\mathbf{X}_i,
  \mathbf{H}_{E,P}),
\end{align}
\begin{align}\label{eq:t3}
&h(\mathbf{X}_i|\mathbf{X}_j,\mathbf{Y}_j^{(2)},
  \mathcal{C})=h(\mathbf{X}_i|\mathbf{X}_j,\mathbf{Y}_j^{(2)},
  \mathbf{H}_j,\mathbf{H}_{E,P},\mathbf{Y}_E^{(2)})\notag\\
  &=h(\mathbf{Y}_j^{(2)},\mathbf{Y}_E^{(2)}|\mathbf{X}_i,\mathbf{X}_j,
  \mathbf{H}_j,\mathbf{H}_{E,P})+h(\mathbf{X}_i|\mathbf{X}_j,
  \mathbf{H}_j,\mathbf{H}_{E,P})\notag\\
  &\,\,-h(\mathbf{Y}_j^{(2)},\mathbf{Y}_E^{(2)}|\mathbf{X}_j,
  \mathbf{H}_j,\mathbf{H}_{E,P})\notag\\
  &=h(\mathbf{Y}_j^{(2)},\mathbf{Y}_E^{(2)}|\mathbf{X}_i,\mathbf{X}_j,
  \mathbf{H}_j,\mathbf{H}_{E,P})+h(\mathbf{X}_i)\notag\\
  &\,\,-h(\mathbf{Y}_j^{(2)},\mathbf{Y}_E^{(2)}|\mathbf{X}_j,
  \mathbf{H}_j,\mathbf{H}_{E,P})
\end{align}
\begin{align}\label{eq:t4}
&h(\mathbf{Y}_i^{(2)}|\mathbf{X}_i,\mathbf{X}_j,\mathbf{Y}_j^{(2)},
  \mathcal{C})\notag\\
  &=h(\mathbf{Y}_i^{(2)}|\mathbf{X}_i,\mathbf{X}_j,\mathbf{Y}_j^{(2)},
  \mathbf{H}_i,\mathbf{H}_j,\mathbf{H}_{E,P},\mathbf{Y}_E^{(2)})\notag\\
  &=h(\mathbf{Y}_i^{(2)},\mathbf{Y}_j^{(2)},\mathbf{Y}_E^{(2)}|\mathbf{X}_i,\mathbf{X}_j,
  \mathbf{H}_i,\mathbf{H}_j,\mathbf{H}_{E,P})\notag\\
  &\,\,-h(\mathbf{Y}_j^{(2)},\mathbf{Y}_E^{(2)}|\mathbf{X}_i,\mathbf{X}_j,
  \mathbf{H}_i,\mathbf{H}_j,\mathbf{H}_{E,P})\notag\\
  &=h(\mathbf{Y}_i^{(2)},\mathbf{Y}_j^{(2)},\mathbf{Y}_E^{(2)}|\mathbf{X}_i,\mathbf{X}_j,
  \mathbf{H}_i,\mathbf{H}_j,\mathbf{H}_{E,P})\notag\\
  &\,\,-h(\mathbf{Y}_j^{(2)},\mathbf{Y}_E^{(2)}|\mathbf{X}_i,\mathbf{X}_j,
  \mathbf{H}_j,\mathbf{H}_{E,P}).
\end{align}
Using \eqref{eq:t1}-\eqref{eq:t4} into \eqref{eq:upper_bound3} yields
\begin{align}\label{eq:upper_bound4}
  &C_{\texttt{key},i,j,\texttt{up}}^{(2)}\approx
  -h(\mathbf{Y}_E^{(2)}|\mathbf{H}_{E,P})\notag\\
  &\,\,+h(\mathbf{Y}_i^{(2)},\mathbf{Y}_E^{(2)}|\mathbf{X}_i,
  \mathbf{H}_i,\mathbf{H}_{E,P})\notag\\
  &\,\,+h(\mathbf{Y}_j^{(2)},\mathbf{Y}_E^{(2)}|\mathbf{X}_j,
  \mathbf{H}_j,\mathbf{H}_{E,P})\notag\\
  &\,\,-h(\mathbf{Y}_i^{(2)},\mathbf{Y}_j^{(2)},\mathbf{Y}_E^{(2)}|\mathbf{X}_i,\mathbf{X}_j,
  \mathbf{H}_i,\mathbf{H}_j,\mathbf{H}_{E,P}).
\end{align}
The DoF of the first term in \eqref{eq:upper_bound4} is given by \eqref{eq:DoFhYEHEP} in section \ref{sec:hYEHEP}, and the DoFs of the second and third terms follow \eqref{eq:DoFhYiYEXiHiHEP} in section \ref{sec:hYiYEXiHiHEP}.
Next we only need to analyze the fourth term in \eqref{eq:upper_bound4}.

  \subsubsection{Analysis of $h(\mathbf{Y}_i^{(2)},\mathbf{Y}_j^{(2)},\mathbf{Y}_E^{(2)}|\mathbf{X}_i,\mathbf{X}_j,
  \mathbf{H}_i,\mathbf{H}_j,\mathbf{H}_{E,P})$}
  Without loss of generality, let us consider  $h(\mathbf{Y}_1^{(2)},\mathbf{Y}_2^{(2)},\mathbf{Y}_E^{(2)}|\mathcal{C}'_2)$ with $\mathcal{C}'_2 = \{\mathbf{X}_1,\mathbf{X}_2,
  \mathbf{H}_1,\mathbf{H}_2,\mathbf{H}_{E,P}\}$.

  Note that $\mathbf{Y}_1^{(2)}=\sigma(\mathbf{H}_{1,2}\mathbf{X}_2+\mathbf{H}_{1,3}\mathbf{X}_3+\cdots+
  \mathbf{H}_{1,M}\mathbf{X}_M)+\mathbf{W}_1^{(2)}$ where the first term $\mathbf{H}_{1,2}\mathbf{X}_2$ is constant given $\mathcal{C}'_2$.
  It follows that
  \begin{align}\label{eq:new1c}
  &\texttt{DoF}(h(\mathbf{Y}_1^{(2)}|\mathcal{C}'_2))
  =\min(N_1,N_T-N_1-N_2)K_2.
  \end{align}
  Let $\mathbf{X}_{(1,2)}$ be the vertical stack of $\mathbf{X}_3,\cdots,\mathbf{X}_M$. Then
  given a high power, $\mathbf{Y}_1^{(2)}$ and $\mathcal{C}'_2$,  we have $\mathbf{X}_{(1,2)}\approx \mathbf{X}_{1,2,0}+\mathbf{N}_{1,2}\mathbf{T}_{1,2}$ where $\mathbf{X}_{1,2,0}\in\mathbb{C}^{(N_T-N_1-N_2)\times K_2}$ and $\mathbf{N}_{1,2}\in\mathbb{C}^{(N_T-N_1-N_2)\times (N_T-2N_1-N_2)^+}$ are constant while $\mathbf{T}_{1,2}\in\mathbb{C}^{(N_T-2N_1-N_2)^+\times K_2}$ is arbitrary. The column span of $\mathbf{N}_{1,2}$ is the right null space of $[\mathbf{H}_{1,3},\cdots,\mathbf{H}_{1,M}]$. It follows that
  \begin{align}\label{eq:new2c}
  &\texttt{DoF}(h(\mathbf{Y}_2^{(2)}|\mathbf{Y}_1^{(2)},\mathcal{C}'_2))\notag\\
  &
  =\min(N_2,(N_T-2N_1-N_2)^+)K_2.
  \end{align}
  Given a high power, $\mathbf{Y}_1^{(2)}$, $\mathbf{Y}_2^{(2)}$ and $\mathcal{C}'_2$, we have
  $\mathbf{X}_{(1,2)}\approx \mathbf{\bar X}_{1,2,0}+\mathbf{\bar N}_{1,2}\mathbf{\bar T}_{1,2}$ where $\mathbf{\bar X}_{1,2,0}\in\mathbb{C}^{(N_T-N_1-N_2)\times K_2}$, $\mathbf{\bar N}_{1,2}\in\mathbb{C}^{(N_T-N_1-N_2)\times (N_T-2N_1-2N_2)^+}$ are constant while $\mathbf{\bar T}_{1,2}\in\mathbb{C}^{(N_T-2N_1-2N_2)^+\times K_2}$ is arbitrary. The column span of $\mathbf{\bar N}_{1,2}$ is the right null space of $\left [\begin{array}{ccc}
                                                                       \mathbf{H}_{1,3} & \cdots & \mathbf{H}_{1,M} \\
                                                                       \mathbf{H}_{2,3} & \cdots & \mathbf{H}_{2,M}
                                                                     \end{array}
  \right ]$.

  Recall the $(\alpha,\beta)$-partitions $\mathbf{Y}_E^{(2)}=[\mathbf{Y}_{E,\alpha}^{(2)},\mathbf{Y}_{E,\beta}^{(2)}]$ as in \eqref{eq:alpha_beta}. We know that due to the unknown $\mathbf{H}_{E,P,\perp}$,
  \begin{align}\label{eq:new3c}
    &\texttt{DoF}(h(\mathbf{Y}_{E,\alpha}^{(2)}|\mathbf{Y}_1^{(2)},\mathbf{Y}_2^{(2)},
    \mathcal{C}'_2))=N_E\min(N_{\texttt{min}},K_2).
  \end{align}

  For $K_2>N_{\texttt{min}}$, following a similar analysis leading to \eqref{eq:new3b} and \eqref{eq:new4b}, we have
  \begin{align}\label{eq:new4c}
    &\texttt{DoF}(h(\mathbf{Y}_{E,\beta}^{(2)}|\mathbf{Y}_{E,\alpha}^{(2)},\mathbf{Y}_1^{(2)},
    \mathbf{Y}_2^{(2)},
    \mathcal{C}'_2))\notag\\
    &=\min(N_E,(N_T-2N_1-2N_2)^+)\Delta K_2.
  \end{align}

  Adding up \eqref{eq:new1c}, \eqref{eq:new2c}, \eqref{eq:new3c} and \eqref{eq:new4c}, but with $N_1$ and $N_2$ replaced by $N_i$ and $N_j$, yields
  \begin{align}\label{eq:DoFhYiYjYEXiXj}
    &\texttt{DoF}(h(\mathbf{Y}_i^{(2)},\mathbf{Y}_j^{(2)},\mathbf{Y}_E^{(2)}|\mathbf{X}_i,\mathbf{X}_j,
  \mathbf{H}_i,\mathbf{H}_j,\mathbf{H}_{E,P}))\notag\\
  &=K_2\min(N_i,N_T-N_i-N_j)\notag\\
  &\,\,+K_2\min(N_j,(N_T-2N_i-N_j)^+)\notag\\
  &\,\,+N_E\min(N_{\texttt{min}},K_2)\notag\\
  &\,\,+\Delta K_2\min(N_E,(N_T-2N_i-2N_j)^+).
  \end{align}

\end{document}